 \documentclass[aps,showpacs,amsmath,amssymbd,dvips,showkeys,linenumbers,preprint]{revtex4}

\usepackage{graphicx}
\usepackage{graphics}
\usepackage{color}

\def \be  {\begin{equation}}
\def \ee  {\end{equation}}
\def \ee  {\end{equation}}
\def \bea {\begin{eqnarray}}
\def \eea {\end{eqnarray}}

\newcommand{\nn}{\nonumber}

\begin{document}

\preprint{ECTP-2013-15\hspace*{0.5cm}and\hspace*{0.5cm}WLCAPP-2013-12}
\title{Polyakov linear SU(3) sigma model: features of higher order moments in dense and thermal hadronic medium}

\author{A. ~Tawfik\footnote{http://atawfik.net/}}
\affiliation{Egyptian Center for Theoretical Physics (ECTP), MTI University, 11571 Cairo, Egypt}
\affiliation{World Laboratory for Cosmology And Particle Physics (WLCAPP), Cairo, Egypt}

\author{N. ~Magdy} 
\affiliation{World Laboratory for Cosmology And Particle Physics (WLCAPP), Cairo, Egypt}

\author{A. Diab} 
\affiliation{World Laboratory for Cosmology And Particle Physics (WLCAPP), Cairo, Egypt}

\begin{abstract}
In order to characterize the higher order moments of the particle multiplicity, we implement the linear-sigma model with Polyakov-loop correction. We first studied the critical phenomena and estimated some thermodynamic quantities. Then, we compared all these results with the first--principle lattice QCD calculations. Then, the extensive study of non-normalized four moments is followed by investigating their thermal and density dependences. We repeat this for moments normalized to temperature and chemical potential. The fluctuations of the second order moment is used to estimate the chiral phase--transition. Then, we implement all these in mapping out the chiral phase transition, which shall be compared with the freeze-out parameters estimated from the lattice QCD simulations and the thermal models are compared with the chiral phase--diagram.

\end{abstract}


\maketitle

\section{Introduction}

It is believed that at high temperatures and densities there should be phase transition(s) between combined  nuclear matter and  quark-gluon plasma (QGP), where quarks and gluons are no longer confined inside hadrons \cite{Rischke:2003mt}. The theoretical and experimental studies of QGP still represent a hot topic in high--energy physics. So far, there are many heavy--ion experiments aiming to create this phase of  matter and to study its properties, for example the Relativistic Heavy-Ion Collider (RHIC) and the Large Hadron Collider (LHC). From theoretical point-of-view, there were - apart from Quantum Chromodynamic (QCD) and its numerical simulations - two main first--principle models, the Polyakov Nambu--Jona--Lasinio model (PNJL) \cite{Fukushima:2003fw,Ratti:2005jh,Fukushima:2008wg,Sanjay:2006,Claudia:2007,Abhijit:2010,Abhijitd82:2010,Kanako:2014} and a combination of the chiral linear--sigma model \cite{Gell Mann:1960} with the Polyakov--loop potential (PLSM) \cite{Mao2010,Shankar:2010} or Polyakov quark meson model (PQM) for three quark flavors (two light and one strange quarks) \cite{Schaefer:2008ax,Schaefer:2009ab,Mao:2010,Kahara:2008}.

The linear--sigma model (LSM) was introduced by Gell--Mann and Levy in 1960 \cite{Gell Mann:1960} long time before inverting QCD as the theory of strong interactions. Many studies have been performed on LSM like $\mathcal{O}(4)$ LSM \cite{Gell Mann:1960}, $\mathcal{O}(4)$  LSM at non zero temperature \cite{Lenaghan:1999si, Petropoulos:1998gt} and $U(N_f)_r  \times U(N_f)_l$ LSM for $N_f=2$, $3$ or even $4$ quark flavors \cite{l, Hu:1974qb, Schechter:1975ju, Geddes:1979nd}. In order to obtain reliable results, there is an extension of LSM; PLSM, in which information about the confining glue sector of the theory was included in form of Polyakov--loop potential. The Polyakov--loop potential is to be extracted from pure Yang--Mills lattice simulations \cite{Polyakov:1978vu, Susskind:1979up, Svetitsky:1982gs,Svetitsky:1985ye}. 

There were so far many studies devoted to investigate the phase diagram and the thermodynamics of LSM and even PLSM at different Polyakov--loop potentials with two  \cite{Wambach:2009ee,Kahara:2008} and three quark flavors \cite{Schaefer:2008ax,Mao:2010}. The thermodynamic properties can be evaluated at finite and vanishing chemical potential like pressure, equation of state, speed of sound, specific heat, trace anomaly and even bulk viscosity \cite{Wambach:2009ee,Mao:2010,Kahara:2008,Schaefer:2008ax}. The results with $N_f=2+1$ quark flavors are compared with recent $N_\tau = 8$ lattice QCD data \cite{HotQCD} and with finer lattice spacing \cite{QCDL}. It was found that the QCD phase diagram can be explored \cite{Wambach:2009ee}.

The present work is mainly devoted to characterizing the higher order moments of particle multiplicity \cite{Tawfik:2012si,Tawfik:2013dba} in PLSM. Therefore, we first start with studying its critical phenomena. Then, we estimate some thermodynamic quantities and compare them with the first--principle lattice QCD simulations \cite{HotQCD,QCDL}.  The extensive study of the first four non--normalized moments is followed by investigating their thermal and density dependences. We repeat this for moments normalized to temperature and chemical potential. The logical step that follows is the implementation of all these in mapping out the chiral phase transition, which shall be compared with the freeze-out parameters ~\cite{Tawfik:2012si,Tawfik:2005qn,Tawfik:2004ss} deduced from thermal models \cite{Tawfik:2013bza,SHMUrQM,Karsch:2003vd,Karsch:2003zq,Redlich:2004gp,Tawfik:2004vv,Tawfik:2004sw,Tawfik:2006yq,Tawfik:2010uh,Tawfik:2010pt,Tawfik:2012zz}  and lattice QCD \cite{KarschA,LQCDA}. 

The present paper is organized as follows.  In section \ref{sec:model}, we give details about the approach (PLSM) implemented in estimating the order parameters, some thermodynamic quantities, the higher order moments of the particle multiplicity and the chiral phase diagram. The Polyakov linear--sigma model is summarized in section \ref{subsec:PLSM}. The mean field approximation shall be outlined in section \ref{subsec:mean field}. Section \ref{sec:Results} gives some features of the PLSM. The phase transition including the quark condensates and the order parameters shall be estimated in section \ref{subsec:condensates}.  Various thermodynamic quantities will be calculated in section \ref{subsec:thermo} and compared with the lattice QCD calculations. The first four order moments of the particle multiplicity shall be elaborated in section \ref{sec:higher}. The fluctuations of the second order moment is used to estimate the chiral  phase transition in section \ref{subsec:phasediagram}. We compare these with the experimentally-deduced freeze-out parameters and the corresponding lattice QCD calculations. Section \ref{sec:conclusion} is devoted to conclusions and outlook.

\section{The Approach}
 \label{sec:model}

\subsection{Polyakov Linear-Sigma Model (Quark-Meson Model)} 
\label{subsec:PLSM}
The Lagrangian of LSM with $N_f =3$  quark flavors and $N_c =3$ color degrees of freedom, where the quarks couple to the Polyakov loop dynamics,  was introduced in Ref. \cite{Schaefer:2008ax,Mao:2010},
\begin{eqnarray}
\mathcal{L}=\mathcal{L}_{chiral}-\mathbf{\mathcal{U}}(\phi, \phi^*, T), \label{plsm}
\end{eqnarray}
where the chiral part of the Lagrangian $\mathcal{L}_{chiral}=\mathcal{L}_q+\mathcal{L}_m$ is of $SU(3)_{L}\times SU(3)_{R}$ symmetry  \cite{Lenaghan,Schaefer:2008hk}. The Lagrangian with $N_f =3$ consists of two parts.  The first part gives the fermionic sector, Eq. (\ref{lfermion}) with a flavor-blind Yukawa coupling $g$ of the quarks. The second part stand for the mesonic contribution, Eq. (\ref{lmeson})   
\begin{eqnarray}
\mathcal{L}_q &=& \sum_f \overline{\psi}_f(i\gamma^{\mu}
D_{\mu}-gT_a(\sigma_a+i \gamma_5 \pi_a))\psi_f, \label{lfermion} \\
\mathcal{L}_m &=&
\mathrm{Tr}(\partial_{\mu}\Phi^{\dag}\partial^{\mu}\Phi-m^2
\Phi^{\dag} \Phi)-\lambda_1 [\mathrm{Tr}(\Phi^{\dag} \Phi)]^2 \nonumber\\&& 
-\lambda_2 \mathrm{Tr}(\Phi^{\dag}
\Phi)^2+c[\mathrm{Det}(\Phi)+\mathrm{Det}(\Phi^{\dag})]
+\mathrm{Tr}[H(\Phi+\Phi^{\dag})].  \label{lmeson}
\end{eqnarray}
The summation $\sum_f$ runs over the three flavors ($f=1, 2, 3$ for u-, d-, s-quark). The flavor--blind Yukawa coupling $g$ should couple the quarks to the mesons \cite{blind}. This explains why many authors designate this as the Quark-Meson Model. The coupling of the quarks to the Euclidean gauge field \cite{Polyakov:1978vu,Susskind:1979up} $A_{\mu}=\delta_{\mu 0}A_0$ is given via the covariant derivative $D_{\mu}=\partial_{\mu}-i A_{\mu}$. In Eq. (\ref{lmeson}), $\Phi$ is a complex $3 \times 3$ matrix depending on the $\sigma_a$ and $\pi_a$ \cite{Schaefer:2008hk}, where $\gamma^{\mu}$ are the chiral spinors, $\sigma_a$ are the scalar mesons and $\pi_a$ are the pseudoscalar mesons. 
\begin{eqnarray}
\Phi= T_a \phi _{a} =T_a(\sigma_a+i\pi_a),\label{Phi}
\end{eqnarray}
where $T_a=\lambda_a/2$ with $a = 0, \cdots, 8$ are the nine generators of the $U(3)$ symmetry group and $\lambda_a$ are the eight Gell--Mann matrices \cite{Gell Mann:1960}. The chiral symmetry is  explicitly broken  by $H$  
 \begin{eqnarray}
H = T_a h_a. \label{H}
\end{eqnarray}
$H$ is a $3 \times 3$ matrix with nine parameters, $h_a$. 

When taking into consideration that the spontaneous chiral symmetry breaking takes part in vacuum state, then a finite vacuum expectation value of the fields $\Phi$ and $\bar{\Phi}$ are conjectured to carry the quantum numbers of the vacuum \cite{Gasiorowicz:1969}. As a result, the diagonal components of the explicit symmetry breaking term $h_0$, $h_3$ and $h_8$ should not vanish \cite{Gasiorowicz:1969}. This leads to exact three finite condensates $\bar{\sigma_0}$, $\bar{\sigma_3}$ and $\bar{\sigma_8}$. On the other hand, $\bar{\sigma_3}$ breaks the isospin symmetry $SU(2)$ \cite{Gasiorowicz:1969}. To avoid this situation, we restrict ourselves to SU(3). This, for instance, can be $N_f= 2+1$ \cite{Schaefer:2008hk} flavor symmetry breaking pattern. Correspondingly, two degenerate light (up and down) and one heavy quark flavor (strange) are assumed. Furthermore, the violation of isospin symmetry is neglected. This facilitates the choice of $h_a$ ($h_0 \neq 0$, $h_3=0$ and $h_8 \neq 0$).  Additional to these, five other parameters should be estimated. These are the squared tree level mass of the mesonic fields $m^2$, two possible coupling constants $\lambda_1$  and $\lambda_2$, Yukawa coupling $g$ and a cubic coupling constant $c$. The latter models the axial $U(1)_A$ anomaly of the QCD vacuum. It is more convenient to convert the condensates $\sigma_0$ and $\sigma_8$ into a pure non--strange and strange parts. To this end, an orthogonal basis transformation from the original basis $\bar{\sigma_0}$ and $ \bar{\sigma_8}$ to the non--strange $\sigma_x$ and strange $\sigma_y$ quark flavor basis is required \cite{Kovacs:2006}. 
\bea
\label{sigms}
\left( {\begin{array}{c}
\sigma _x \\
\sigma _y
\end{array}}
\right)=\frac{1}{\sqrt{3}} 
\left({\begin{array}{cc}
\sqrt{2} & 1 \\
1 & -\sqrt{2}
\end{array}}\right) 
\left({ \begin{array}{c}
\sigma _0 \\
\sigma _8
\end{array}}
\right)
\eea

The second term in Eq. (\ref{plsm}), $\mathbf{\mathcal{U}}(\phi, \phi^*, T)$, represents the Polyakov--loop effective potential \cite{Polyakov:1978vu}, which is expressed by using the dynamics of the thermal expectation value of a color traced Wilson loop in the temporal direction  
\bea
\Phi (\vec{x})=\frac{1}{N_c} \langle \mathcal{P}(\vec{x})\rangle ,
\eea
Then, the Polyakov--loop potential and its conjugate read 
\begin{eqnarray}
\phi &=& (\mathrm{Tr}_c \,\mathcal{P})/N_c, \label{phais1}\\ 
\phi^* &=& (\mathrm{Tr}_c\,  \mathcal{P}^{\dag})/N_c, \label{phais2}
\end{eqnarray}
where $\mathcal{P}$ is the Polyakov loop.  This can be represented by a matrix in the color space \cite{Polyakov:1978vu} 
\begin{eqnarray}
 \mathcal{P}(\vec{x})=\mathcal{P}\mathrm{exp}\left[i\int_0^{\beta}d \tau A_4(\vec{x}, \tau)\right],\label{loop}
\end{eqnarray}
where $\beta=1/T$ is the inverse temperature and $A_4 = i A^0$ is called Polyakov gauge \cite{Polyakov:1978vu,Susskind:1979up}.

The Polyakov loop matrix can be represented  as a diagonal representation \cite{Fukushima:2003fw}. The coupling between the Polyakov loop and the quarks is given by the covariant derivative of $D_{\mu}=\partial_{\mu}-i A_{\mu}$  in PLSM Lagrangian, Eq. (\ref{plsm}). $A_{\mu}=\delta_{\mu 0} A_0$  is restricted to the chiral limit. It is apparent that the PLSM Lagrangian, Eq. (\ref{plsm}), is invariant under the chiral flavor group. This is similar to the original QCD Lagrangian \cite{Ratti:2005jh,Roessner:2007,Fukushima:2008wg}. In order to reproduce the thermodynamic behavior of the Polyakov loop for pure gauge case, we use a temperature--dependent potential $U(\phi, \phi^{*},T)$. This should agree with the lattice QCD simulations and have $Z(3)$ center symmetry as that of the pure gauge QCD Lagrangian \cite{Ratti:2005jh,Schaefer:2007d}. In case of no quarks, then $\phi = \phi^{*}$ and the Polyakov loop is considered as an order parameter for the deconfinement phase--transition  \cite{Ratti:2005jh,Schaefer:2007d}. In the present work, we use $U(\phi, \phi^{*},T)$ as a polynomial expansion in $\phi$ and $\phi^{*}$ \cite{Ratti:2005jh,Roessner:2007,Schaefer:2007d,Fukushima:2008wg}
 \begin{eqnarray}
\frac{\mathbf{\mathcal{U}}(\phi, \phi^*, T)}{T^4}=-\frac{b_2(T)}{2}|\phi|^2-\frac{b_3
}{6}(\phi^3+\phi^{*3})+\frac{b_4}{4}(|\phi|^2)^2, \label{Uloop}
\end{eqnarray}
where 
\begin{eqnarray}
b_2(T)=a_0+a_1\left(\frac{T_0}{T}\right)+a_2\left(\frac{T_0}{T}\right)^2+a_3\left(\frac{T_0}{T}\right)^3. 
\end{eqnarray}
 
In order to reproduce pure gauge QCD thermodynamics and the behavior of the Polyakov loop as a function of temperature, we use the parameters listed out in Tab. \ref{parameter}
\begin{table}[hbt]
\begin{center}

\begin{tabular}{c}
\hline  
$ a_0=6. 75 $,\qquad $ a_1=-1. 95 $, \qquad $ a_2=2. 625 $, \qquad
 $ a_3=-7. 44 $ \\  $ b_3 = 0.75 $ \qquad $b_4=7.5 $   \\ 
\hline 
\end{tabular} 
\caption{The potential parameters are adjusted to the pure gauge lattice data such that the equation of state and the Polyakov loop expectation values are reproduced \cite{Ratti:2005jh}. \label{parameter}}
\end{center}
\end{table}
For a much better agreement with the lattice QCD results, the deconfinement temperature $T_0$  in pure gauge sector is fixed at $270~$MeV. 

 \subsection{The Mean Field  Approximation}
\label{subsec:mean field} 

To calculate the grand potential in the mean field approximation,  we start from the partition function. In thermal equilibrium, the grand partition function can be defined by using a path integral over the quark, antiquark and meson field. 
\begin{eqnarray}
\mathcal{Z}&=& \mathrm{Tr \,exp}[-(\hat{\mathcal{H}}-\sum_{f=u, d, s}
\mu_f \hat{\mathcal{N}}_f)/T] \nonumber\\
&=& \int\prod_a \mathcal{D} \sigma_a \mathcal{D} \pi_a \int
\mathcal{D}\psi \mathcal{D} \bar{\psi} \mathrm{exp} \left[ \int_x
(\mathcal{L}+\sum_{f=u, d, s} \mu_f \bar{\psi}_f \gamma^0 \psi_f )
\right],
\end{eqnarray}
where $\int_x\equiv i \int^{1/T}_0 dt \int_V d^3x$ and $V$ is the volume of the system. $\mu_f$ is the chemical potential for quark flavors $f=(u, d, s)$. We consider symmetric quark matter and define a uniform blind chemical potential $\mu_f \equiv \mu_{u, d} = \mu_s$. Then, we evaluate the partition function in the mean field approximation  \cite{Schaefer:2007c,Schaefer:2008hk,blind}. The meson fields can be replaced by their expectation values $\bar{\sigma_x}$ and $\bar{\sigma_y}$ in the action \cite{Kapusta:2006pm,Mao:2010}. We can use the standard methods \cite{Kapusta:2006pm} in order to calculate the integration over the fermions yields. Then, the effective potential for the mesons can be deduced. Now, we define the thermodynamic potential density 
\begin{eqnarray}
\Omega(T, \mu)=\frac{-T \mathrm{ln}
\mathcal{Z}}{V}=U(\sigma_x, \sigma_y)+\mathbf{\mathcal{U}}(\phi, \phi^*, T)+\Omega_{\bar{\psi}
\psi}. \label{potential}
\end{eqnarray}

The quarks and antiquarks contributions to the potential were introduced in Ref.  \cite{Kapusta:2006pm,Mao:2010}
\begin{eqnarray} \label{qqpotio}
\Omega_{\bar{\psi} \psi} &=& -2 T N \int_0^{\infty} \frac{d^3\vec{p}}{(2 \pi)^3} \left\{ \ln \left[ 1+3(\phi+\phi^* e^{-(E-\mu)/T})\times e^{-(E-\mu)/T}+e^{-3 (E-\mu)/T}\right] \right. \nonumber \\ 
&& \hspace*{30mm} \left.  +\ln \left[ 1+3(\phi^*+\phi e^{-(E+\mu)/T})\times e^{-(E+\mu)/T}+e^{-3 (E+\mu)/T}\right] \right\}, 
\end{eqnarray}
where $N$ gives the number of the quark flavors, $E=\sqrt{\vec{p}^2+m^2}$ is the energy of the valence quark and antiquark energy for light and strange quark. Also, the light quark sector decouples from the strange quark sector \cite{Kovacs:2006}. Assuming degenerate light quarks, i.e. $q\equiv u, d$, then the masses can be simplified as follows.
\begin{eqnarray}
m_q &=& g \frac{\sigma_x}{2}, \label{qmass} \\
m_s &=& g \frac{\sigma_y}{\sqrt{2}}.  \label{sqmass}
\end{eqnarray} 

The purely mesonic potential is given as
 \begin{eqnarray}
U(\sigma_x, \sigma_y) &=& - h_x \sigma_x - h_y \sigma_y + \frac{m^2}{2} (\sigma^2_x+\sigma^2_y) - \frac{c}{2\sqrt{2}} \sigma^2_x \sigma_y \nonumber \\&& + \frac{\lambda_1}{2} \sigma^2_x \sigma^2_y +\frac{1}{8} (2 \lambda_1
+\lambda_2)\sigma^4_x + \frac{1}{4} (\lambda_1+\lambda_2)\sigma^4_y. \label{Upotio}
\end{eqnarray}
We notice that the sum in Eqs. (\ref{Uloop}), (\ref{qqpotio}) and (\ref{Upotio}) represent the thermodynamic potential density as given in Eq. (\ref{potential}), which has seven parameters $m^2, h_x, h_y, \lambda_1, \lambda_2, c$ and $ g$ two unknown condensates $\sigma_x$ and $ \sigma_y$ and an order parameter for the deconfinement $\phi$ and $\phi^*$. The six parameters $m^2, h_x, h_y, \lambda_1, \lambda_2 $ and $c$  are fixed in vacuum by six experimentally known quantities \cite{Schaefer:2008hk}. In order to evaluate the unknown parameters $\sigma_x$, $ \sigma_y$, $\phi$ and $\phi^*$, we minimize the thermodynamic potential, Eq. (\ref{potential}), with respect to $\sigma_x$, $ \sigma_y$, $\phi$ and $\phi^*$. Doing this, we obtain a set of four equations of motion
\begin{eqnarray}\label{cond1}
\left.\frac{\partial \Omega}{\partial \sigma_x}= \frac{\partial
\Omega}{\partial \sigma_y}= \frac{\partial \Omega}{\partial
\phi}= \frac{\partial \Omega}{\partial \phi^*}\right|_{min} =0,
\end{eqnarray}
meaning that $\sigma_x=\bar{\sigma_x}$, $\sigma_y=\bar{\sigma_y}$, $\phi=\bar{\phi}$ and $\phi^*=\bar{\phi^*}$ are the global minimum. 

\section{The Results}
 \label{sec:Results}

\subsection{Phase Transition: Quark Condensates and Order Parameters}
\label{subsec:condensates}

In this section, we study the dependence of the chiral condensates, $\sigma_x$ and $\sigma_y$, and the order parameters, $\phi$ and $\phi^*$, on the temperature and the chemical potential.  Using the minimization conditions given in Eq. (\ref{cond1}), we obtain the dependence of the potential on the three parameters, the temperature, the chemical potential and the minimization parameter. The latter assures a minimum potential, as well. Apparently, this depends on the temperature and the chemical potential. Additionally, we have other four parameters, $\sigma_x$, $\sigma_y$, $\phi $ and $\phi^*$. Therefore, the minimization step should be repeated for each of these parameters, while the other parameters should remain fixed, i.e. {\it global minimum of other parameters}. Repeating this process, we get each parameter as a function of the temperature and the chemical potential. 

\begin{figure}[htb]
\includegraphics[width=6cm,angle=-90]{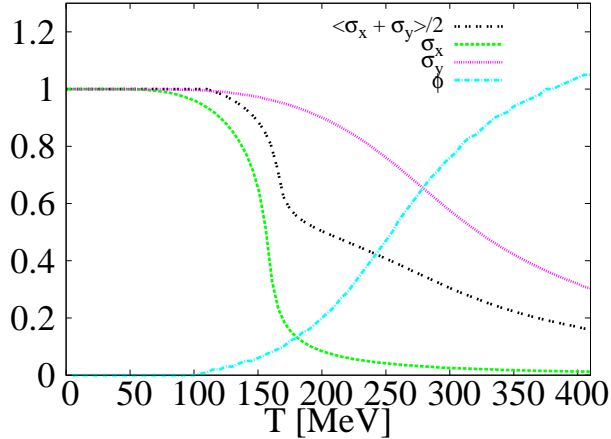}
\caption{(Color online) The normalized chiral condensates $\sigma _{x}$ and $ \sigma_{y}$ (dashed and dotted curves, respectively) and the Polyakov loops $\phi$ and $\phi ^{*}$ (dot--dashed curve) are given as functions of temperature at vanishing chemical potential. At $ \mu=0~$MeV, the two Polyakov loops are identical, i. e. $\phi$=$\phi^{*}$. The double--dotted curve represents the averaged chiral condensate including light and strange quark flavors. }
\label{fig:Ttrans} 
\end{figure}

In Fig. \ref{fig:Ttrans}, the thermal evolution of the normalized chiral condensates, $\sigma$'s, of light (dashed curve) and strange quark flavors (dotted curve) and the order parameter from the two Polyakov loops, $\phi$ and $\phi^{*}$, (dot--dashed curve) are analysed. The latter quantities are conjectured to give a reliable indication for the non--strange and the strange critical temperatures. The non--strange phase transition is related to $\sigma_x$, while $\sigma_y$ reflects the strange phase--transition. The dash--dotted curve represents the averaged chiral condensation, in which light and strange quark flavors are included. The chiral condensates are normalized to their zero--temperature values at vanishing temperature, $\sigma_{x_{0}}=92.4~$MeV and $\sigma_{y_{0}}=94.5~$MeV for light and strange quarks, respectively \cite{Schaefer:2008hk,Mao:2010}. This normalization is also valid at vanishing chemical potential. At $\mu=0~$MeV, the two Polyakov loops are identical, i. e. $\phi$=$\phi^{*}$. The point, at which the order parameter intersects the curve of corresponding chiral condensate is taken as the critical temperature. Accordingly, we roughly estimate $T_{c,\, q}\sim 175$,  $T_{c,\, s}\sim 275$ and $T_{c,\, q+s}\sim 240~$MeV.

\begin{figure}[htb]
\centering{
\includegraphics[width=5.cm,angle=-90]{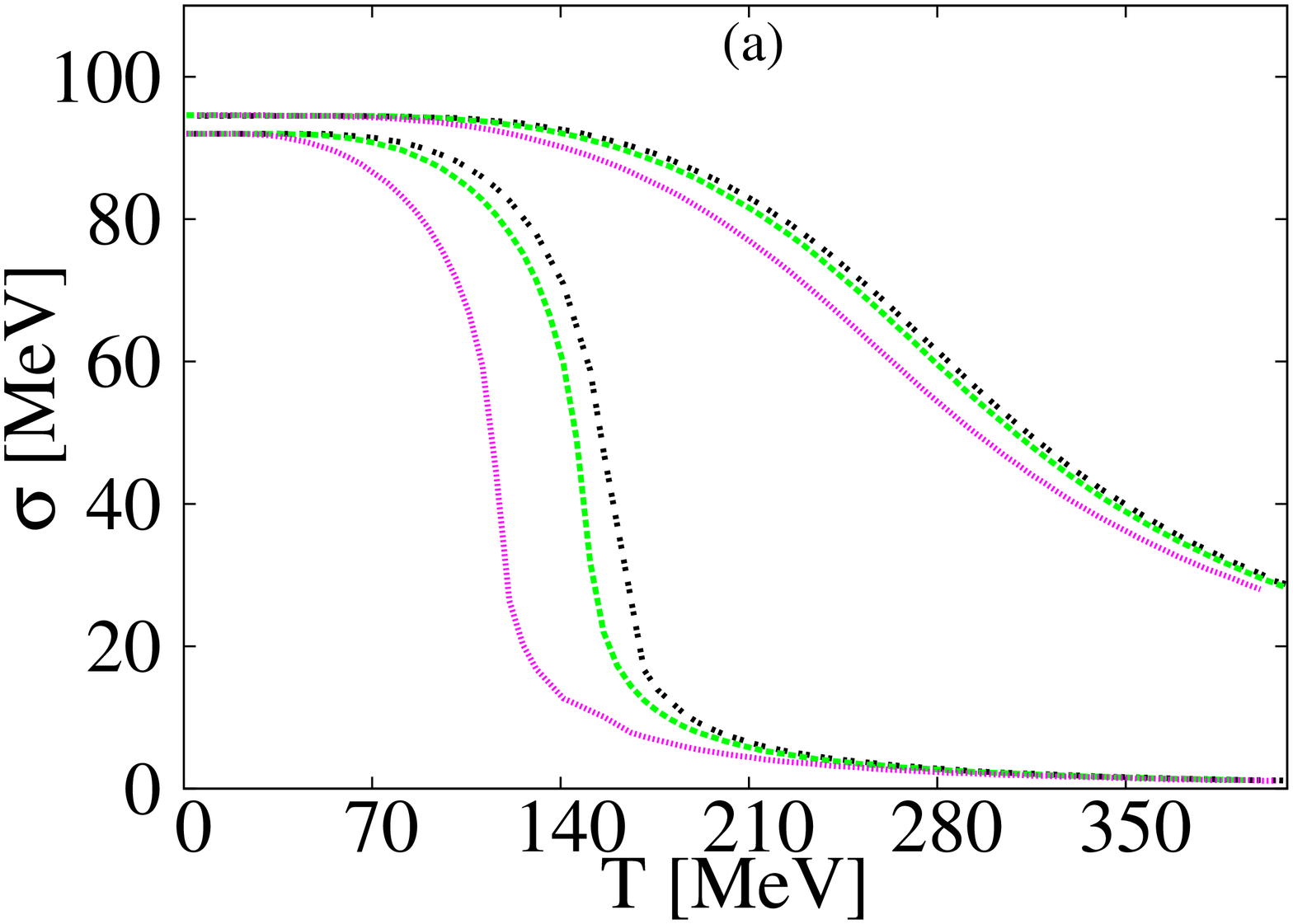}
\includegraphics[width=5.cm,angle=-90]{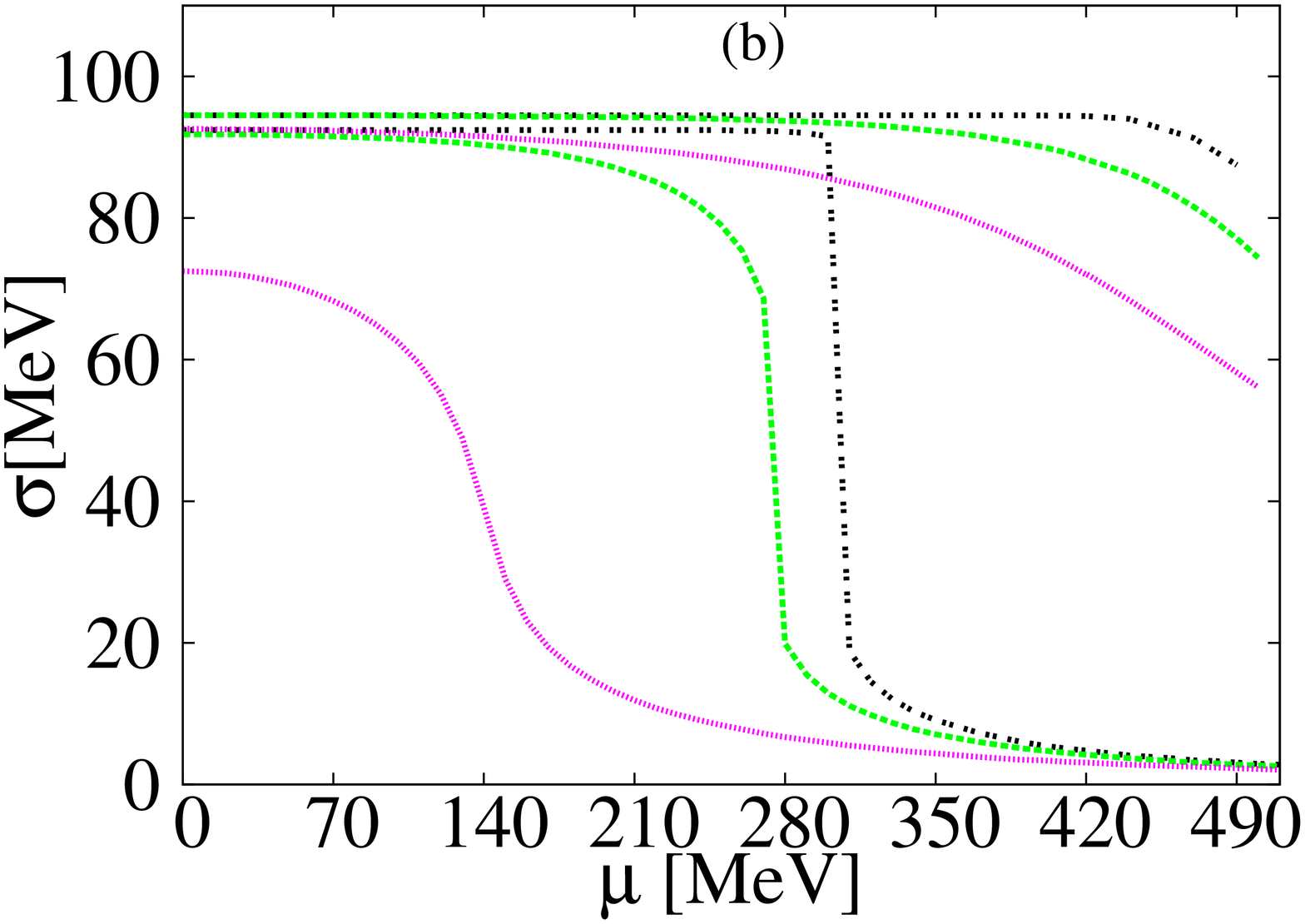}
\caption{(Color online) Light chiral condensates, $\sigma_x$ and $\sigma_y$, (left-hand panel) are given as functions of temperature at different chemical potentials, $\mu\, =\,0$ MeV (dotted curve), $100$ MeV (dashed curve) and $200~$ MeV (double--dotted curve). Light chiral condensates, $\sigma_x$ and $\sigma_y$, (right--hand panel) are given as functions of chemical potential  at different temperatures $10$ MeV (dotted curve), $70$ MeV (dashed curve) and $140~$MeV (double--dotted curve). \label{fig:2}} 
}
\end{figure}

When the chemical potential $\mu$ is switched on (when the system of interest in emerged in a dense medium), the thermal evolution of the non--normalized chiral condensates $\sigma$'s is given in the left--hand panel of Fig. \ref{fig:2}.  We find that increasing chemical potential decreases the values of the chiral condensates, $\sigma_{x}$ and $\sigma_{y}$. So far, the vacuum expectation values of the two condensates are measured at vanishing chemical potential. We also draw non--normalized chiral condensates. The values of $\sigma_{x}$ is smaller than that of $\sigma_{y}$. It is apparent that $\sigma_{x}$ decreases much faster than $\sigma_{y}$.  The thermal behavior of the two chiral condensates is conjectured to characterize the phase transition. Below $T_c$, there is no any change in the quark masses since the condensates is conjectured to remain constants. With increasing $T$, the system moves to a region, in which the quarks lose their masses and the broken chiral symmetry is conjectured to be restored. 

In the right--hand panel of Fig. \ref{fig:2}, the two chiral condensate are given as functions of the chemical potential at different temperatures. Also here, increasing $T$ decreases the values of $\sigma_{x}$ and $\sigma_{y}$. The decrease of both quantities in dense medium is relatively sudden, especially at low temperature. This is related to the chiral phase--transition at large density and low temperature (very near to the abscissa of the QCD phase diagram \cite{Tawfik:2004sw}). The results for $\sigma_x$ refer to a prompt phase--transition, which likely would be characterized as a first order. The related quark chemical potential is $\sim\,310~$MeV. Such a prompt change seems to decrease with decreasing $\mu$. At $\sim\,200~$MeV, a prompt change between confined and deconfined phases disappears. This would be interpreted as a {\it smooth} phase transition know as cross--over \cite{lQCDco}.  Furthermore, the results for $\sigma_y$ are corresponding to larger chemical potential relative to that of $\sigma_x$. Here, the dependence on $\mu$ is smother than that of  $\sigma_x$. Before, we draw any conclusion out of this behavior,  further analysis would help in conducting further verification.

\begin{figure}[htb]
\centering{
\includegraphics[width=5.cm,angle=-90]{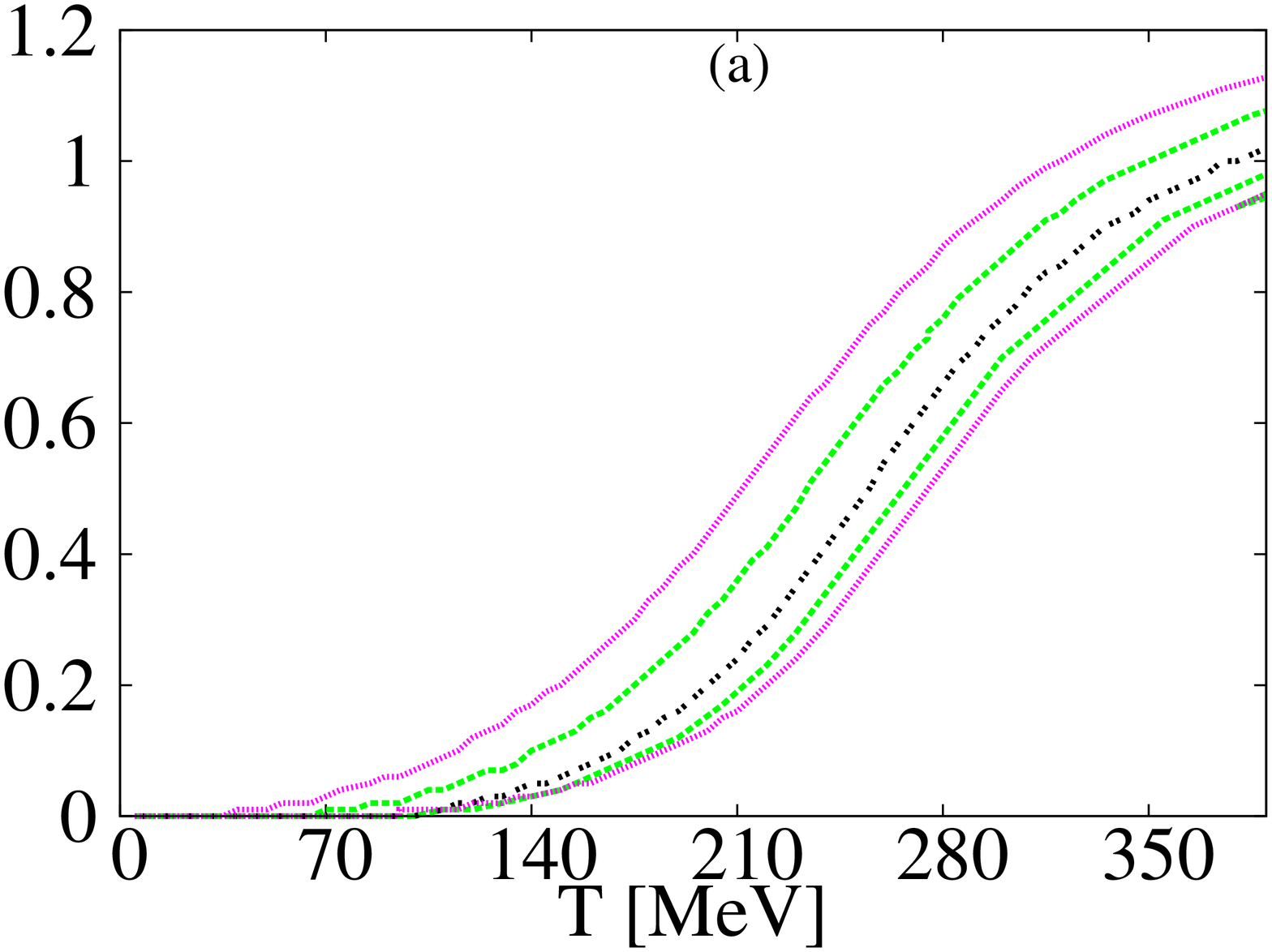}
\includegraphics[width=5.cm,angle=-90]{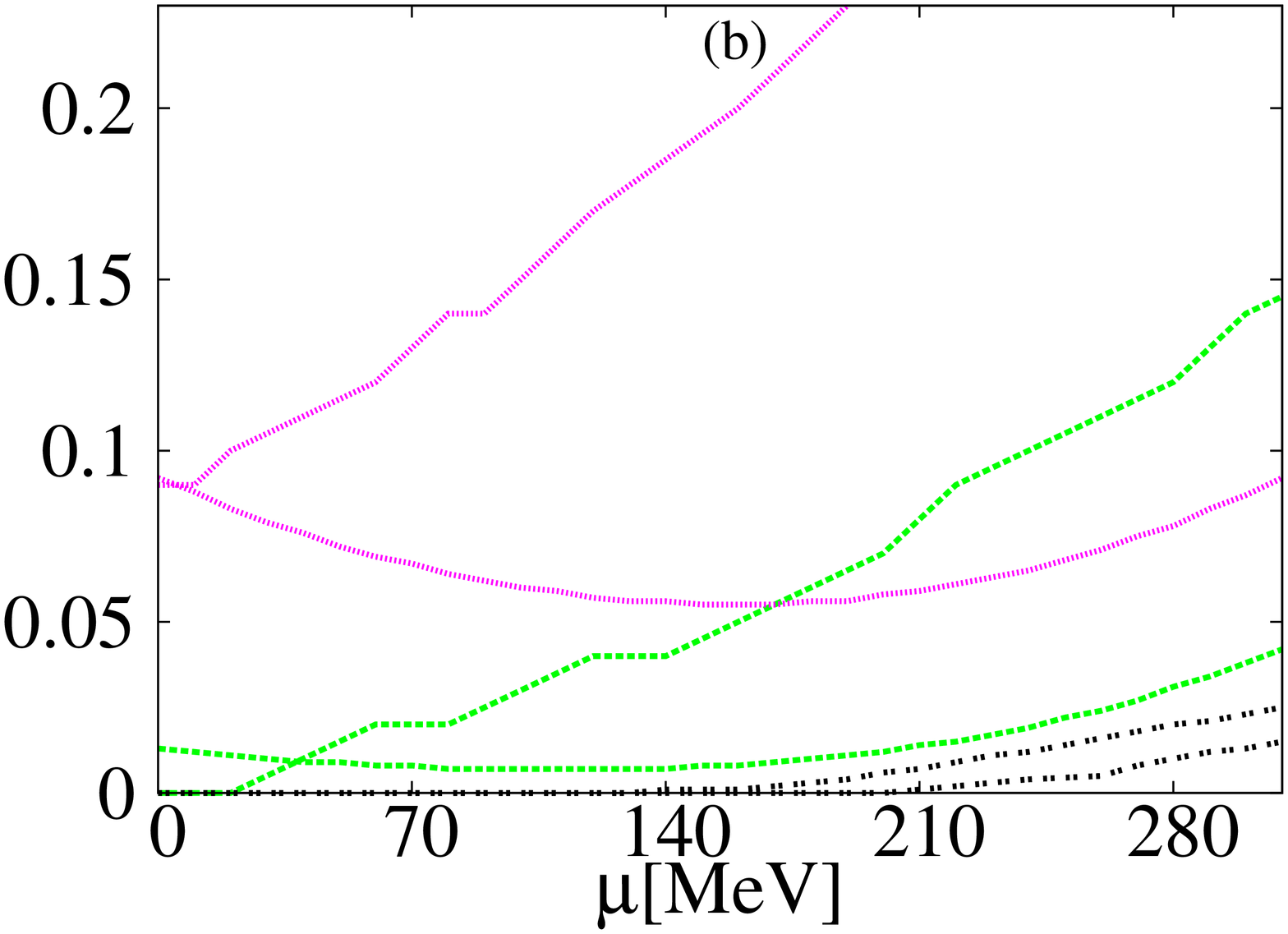}
\caption{(Color online) Left--hand panel: the order parameters $\phi$ and $\phi^*$ are given as functions of temperature $T$ at different chemical potentials. Double--dotted curves represent the results at vanishing $\mu$. Dashed and dotted curves give both order parameters at $T=100~$ and $T=200~$MeV, respectively.  The right--hand panel shows the dependence of the order parameters on the chemical potential at different values of $T$.  \label{fig:3} }
}
\end{figure}

Another smooth dependence is shown in Fig. \ref{fig:3}. In the left--hand panel, the Polyakov loops, $\phi$ and $\phi ^{*}$, are given as functions of the temperature at different chemical potentials. Increasing $\mu$ increases the values of $\phi $, but simultaneously decreases that of $\phi^{*}$. Here, the order parameters $\phi$ and $\phi^{*}$ are related to the deconfinement phase--transition. At $\mu=200~$MeV (dotted curve), the thermal evolution of $\phi$ seems to be very smooth. The double--dotted curves represent the results at vanishing $\mu$. The dashed curves give the values of both order parameters at $\mu=100~$MeV.  In the right--hand panel, the dependence of $\phi$ and $ \phi ^{*}$ on $\mu$ is studied at different values of $T$. We notice that $\phi$ has a stronger dependence than $\phi^{*}$. Also, the slope of $\phi(\mu)$ is positive, while $\phi^{*}(\mu)$ slowly decreases and then slowly increase with increasing $\mu$. Both quantities intersect at a characteristic value of $\mu$ depending on $T$. 

\begin{figure}[htb]
\centering{
\includegraphics[width=6.cm,angle=-90]{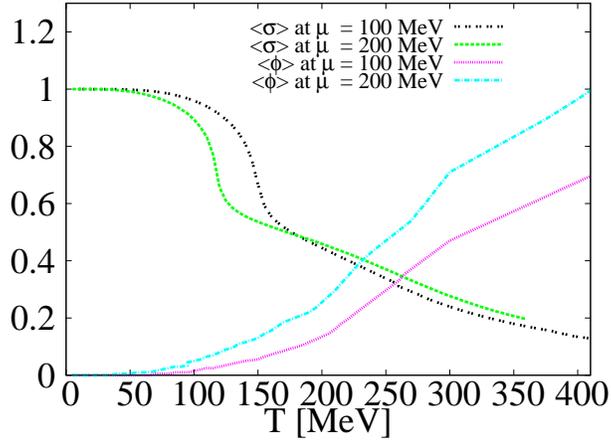}
\caption{(Color online) The thermal behavior of $\langle\sigma\rangle=(\sigma_x+\sigma_y)/2$ and $\langle\phi\rangle=(\phi+\phi^*)/2$ at different values of the chemical potential $\mu$. \label{fig:4}}
}
\end{figure}

For completeness, we show in Fig. \ref{fig:4} the thermal behavior of $\langle\sigma\rangle=(\sigma_x+\sigma_y)/2$ and $\langle\phi\rangle=(\phi+\phi^*)/2$ at different chemical potentials (see Fig. \ref{fig:Ttrans}). It seems that both order quantities intersect at different $\mu$'s. Further studies should favor or disfavor the observation of the critical temperature.

Furthermore, we want to recall that many authors used to calculate the  critical values from the peak values of the temperature variation of $\phi$ and $\sigma$. In section \ref{subsec:thermo}, we discuss the significance of the intersection method, which introduce in the present work, and compare it with the other method.

\subsection{Thermodynamic Quantities}
\label{subsec:thermo}

\begin{figure}[htb]
\includegraphics[width=6.cm,angle=-90]{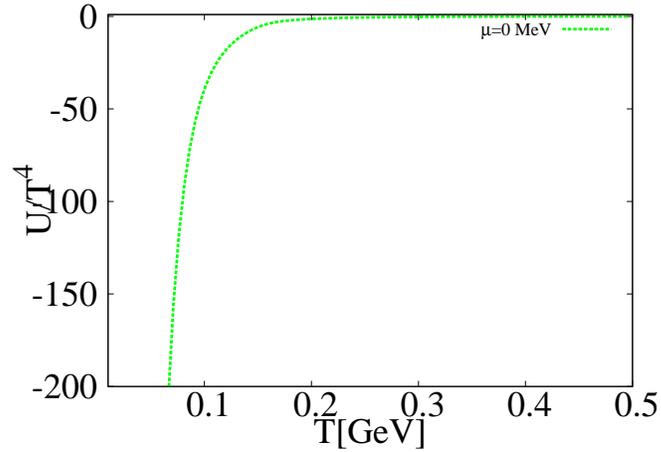}
\caption{(Color online) The thermal evolution of the mesonic potential of LSM is studied at vanishing chemical potential. Accordingly, this part of potential can be excluded, especially at high temperatures.}
\label{fig:LSMpotential} 
\end{figure}

The thermodynamics of LSM and PLSM has been addressed in many studies, such as \cite{Wambach:2009ee,Kahara:2008,Schaefer:2008ax,Mao:2010}. Up to three quark flavors were introduced to PLSM. It was found that the potential can be constructed. The phase diagram of $T$ and $\mu$ can be explored. Also, the thermodynamic properties of the system  like pressure, equation of state,  speed of sound, specific heat, trace anomaly and even bulk viscosity \cite{Wambach:2009ee,Kahara:2008,Schaefer:2008ax,Mao:2010} can be evaluated at vanishing chemical potential $\mu=0$ and eventually compared to the lattice QCD calculations \cite{Mao:2010QCD}. For example, the results from PQM model with $N_f=2+1$ quark flavors  are compared with recent $N_\tau = 8$ lattice calculations \cite{HotQCD}. Another comparison was performed with lattice calculations with finer lattice spacing \cite{QCDL}. Also, the thermodynamic properties of the system at different forms for the Polyakov loop have been evaluated \cite{Wambach:2009ee}. 

In this section, we compare the PLSM results with the lattice QCD calculations \cite{HotQCD,QCDL}. First, we want to estimate the contribution of the purely mesonic potential, Eq. (\ref{Upotio}), would be excluded. At low $T$, this part of potential becomes infinity, but entirely vanishes at high $T$. In Fig. \ref{fig:LSMpotential}, the normalized mesonic potential contribution is given as a function of $T$ at vanishing $\mu$.  Accordingly, this part of potential is only effective at very low temperatures. Therefore, the effective potential, Eq. (\ref{potential}), is simply reduced to  
\begin{eqnarray}
\Omega(T, \mu)=\mathbf{\mathcal{U}}(\phi, \phi^*, T)+\Omega_{\bar{\psi} \psi}, \label{newpotential}
\end{eqnarray}
The pressure density $P$ can obtained  from the grand potential, directly 
 \begin{eqnarray}
P &=& - \Omega(T, \mu).  \label{Pr}
\end{eqnarray}
In the previous sections, we estimated all parameters of the two fields and the two order parameters, as well. Thus, we can now substitute these into Eqs. (\ref{newpotential}) and (\ref{Pr}).

\begin{figure}[htb]
\centering{
\includegraphics[width=5.cm,angle=-90]{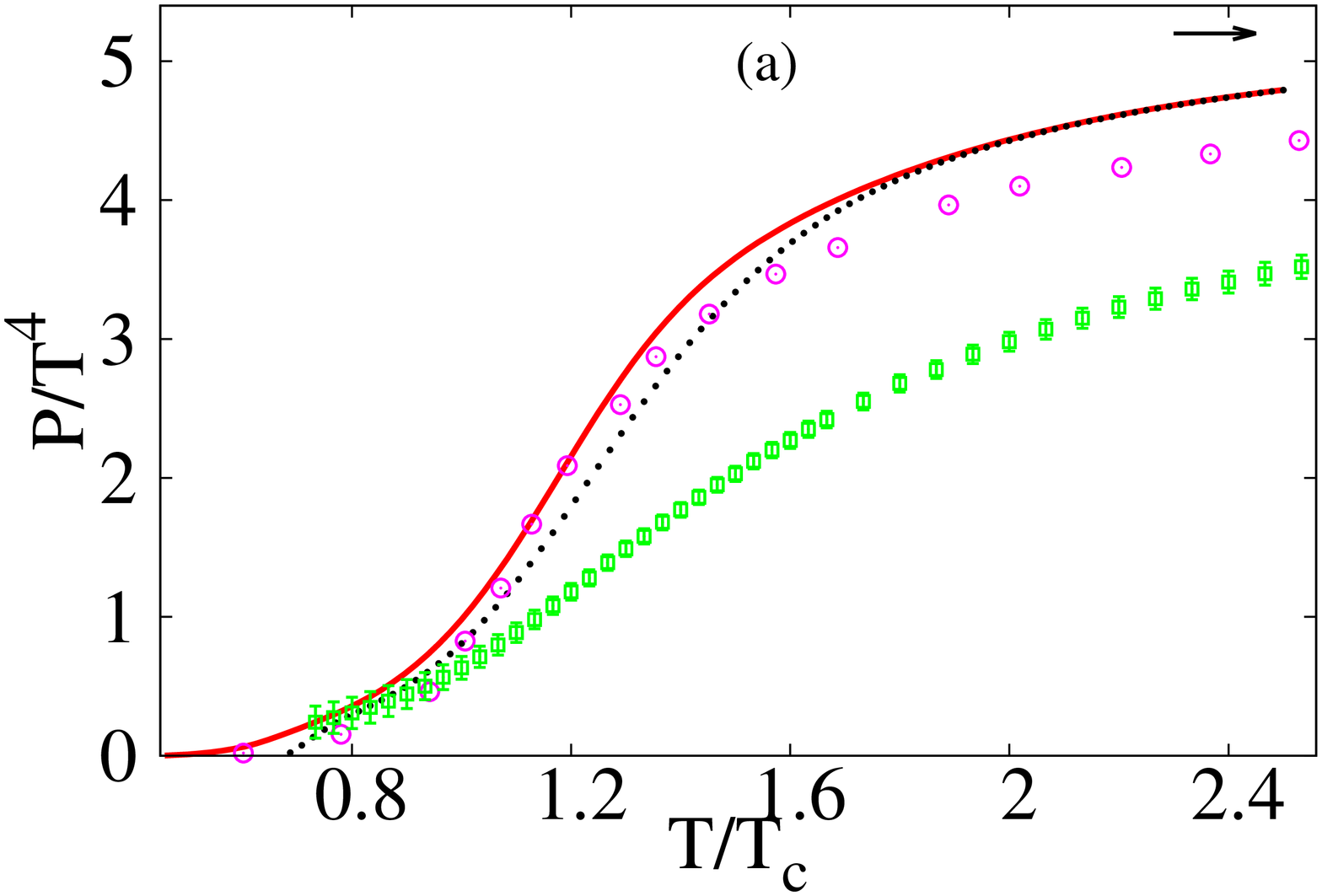}
\includegraphics[width=5.cm,angle=-90]{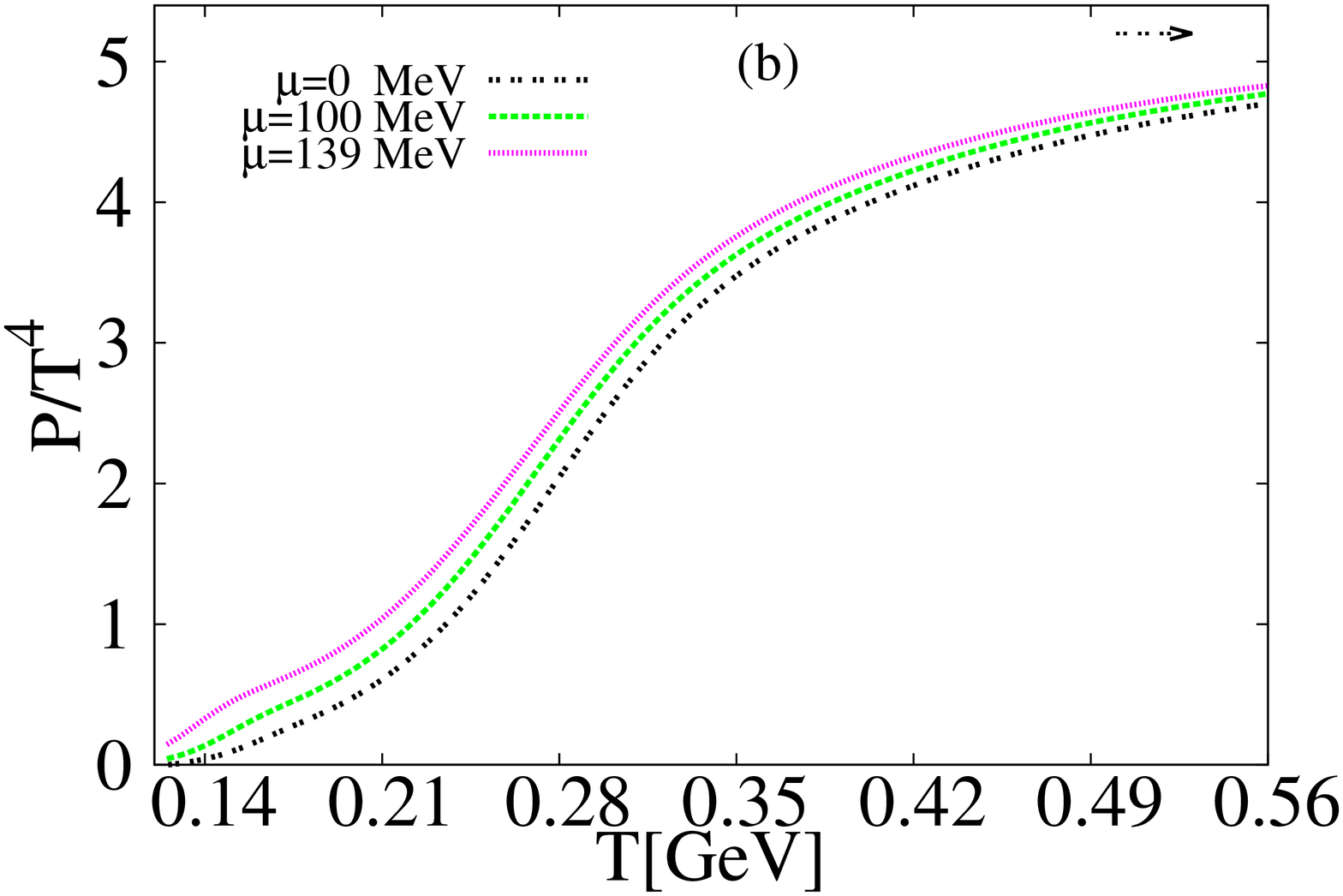}
\caption{(Color online) Left--hand panel shows the thermal behavior of PLSM (curves) and lattice QCD pressure (symbols) at vanishing chemical potential compared with lattice QCD calculations (circles) \cite{HotQCD} (rectangles) \cite{QCDL}. The right--hand panel shows the thermal behavior but at different chemical potentials $\mu=\,0~$MeV (double--dotted curve), $100~$MeV (dashed curve) and $139~$MeV (dotted curve).  \label{fig:pr}}
}
\end{figure}

In the left--hand panel of Fig. \ref{fig:pr}, the PLSM  pressure  is compared with the lattice QCD calculations \cite{HotQCD} (empty circles) \cite{QCDL} (empty rectangles) at vanishing chemical potential. The general $T$--dependence of the pressure is not absence. The pressure increases with $T$ until it gets close to the value of massless gas (Stefan--Boltzmann limit $5.2$). The solid curve represents the results at vanishing temperature chiral condensates, $\sigma_{x\, 0}=92.4~$MeV, $\sigma_{y\, 0}=94.5~$MeV and Yukawa coupling $g=6.5$. Increasing Yukawa coupling to $10.5$ results in the dotted curve. The variation in model parameters has one reason, we need to figure out that this model with its usual parameters can't fit the recent lattice. Both curves fit good the lattice QCD calculations \cite{HotQCD} (circles). It is obvious that the agreement between PLSM and lattice QCD \cite{QCDL} (rectangles) is not convincing, especially above $T_c$. In a forthcoming paper, we present new configurations of LSM so that the most recent lattice simulations \cite{QCDL} (rectangles) turn to be reproducible \cite{TM}, as well. The right-hand panel of Fig. \ref{fig:pr} shows the PLSM pressure as a function of $T$ at different chemical potentials. At finite chemical potentials, the pressure slightly increases.

In another work \cite{TM}, these results shall be confronted to the recent lattice QCD calculations. It is worthwhile to mention that the critical temperature is not universally constant in all these results: \cite{QCDL} and \cite{HotQCD} assume that $T_c\simeq150~$ and $195~$MeV, respectively, while PLSM uses $T_{\chi}=240~$MeV, Fig. \ref{fig:Ttrans}. The changes undertaken in this model have one purpose. We want to figure out how the model parameters should be adjusted to reproduce the recent lattice QCD calculations \cite{QCDL}. First, we made it clear that the {\it original} LSM (without any improvement like Polyakov loop potential) dos not agree with any of the two sets of lattice QCD data. But when adding Polyakov loop potential as done in the present paper, it is possible to simulate excellently the lattice QCD calculations \cite{HotQCD}. On the other hand, this model (PLSM) is not able to simulate the most recent data \cite{QCDL}. The reason would be the fact that the Polyakov loop potential represents the gluonic interaction. The polynomial form for the Polyakov loop potential would enhance this type of interactions and therefore PLSM is not able to reproduce the recent lattice QCD calculations \cite{QCDL}, as well.

The comparison with lattice QCD at finite $\mu$ is not presented here. We wan to report that the lattice QCD pressure at finite $\mu$ is found larger than the PLSM pressure.

\begin{figure}[htb]
\centering{
\includegraphics[width=5.cm,angle=-90]{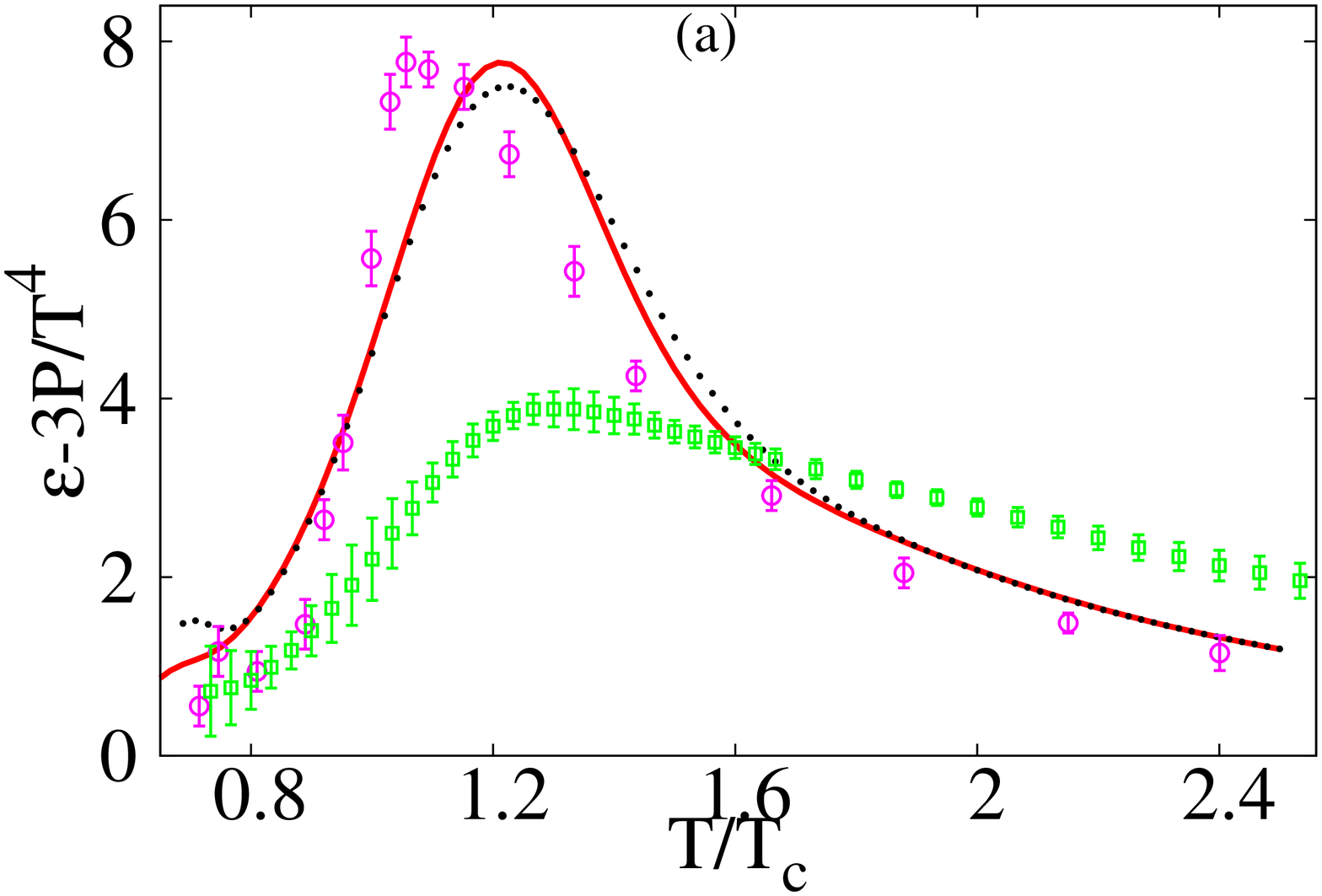}
\includegraphics[width=5.cm,angle=-90]{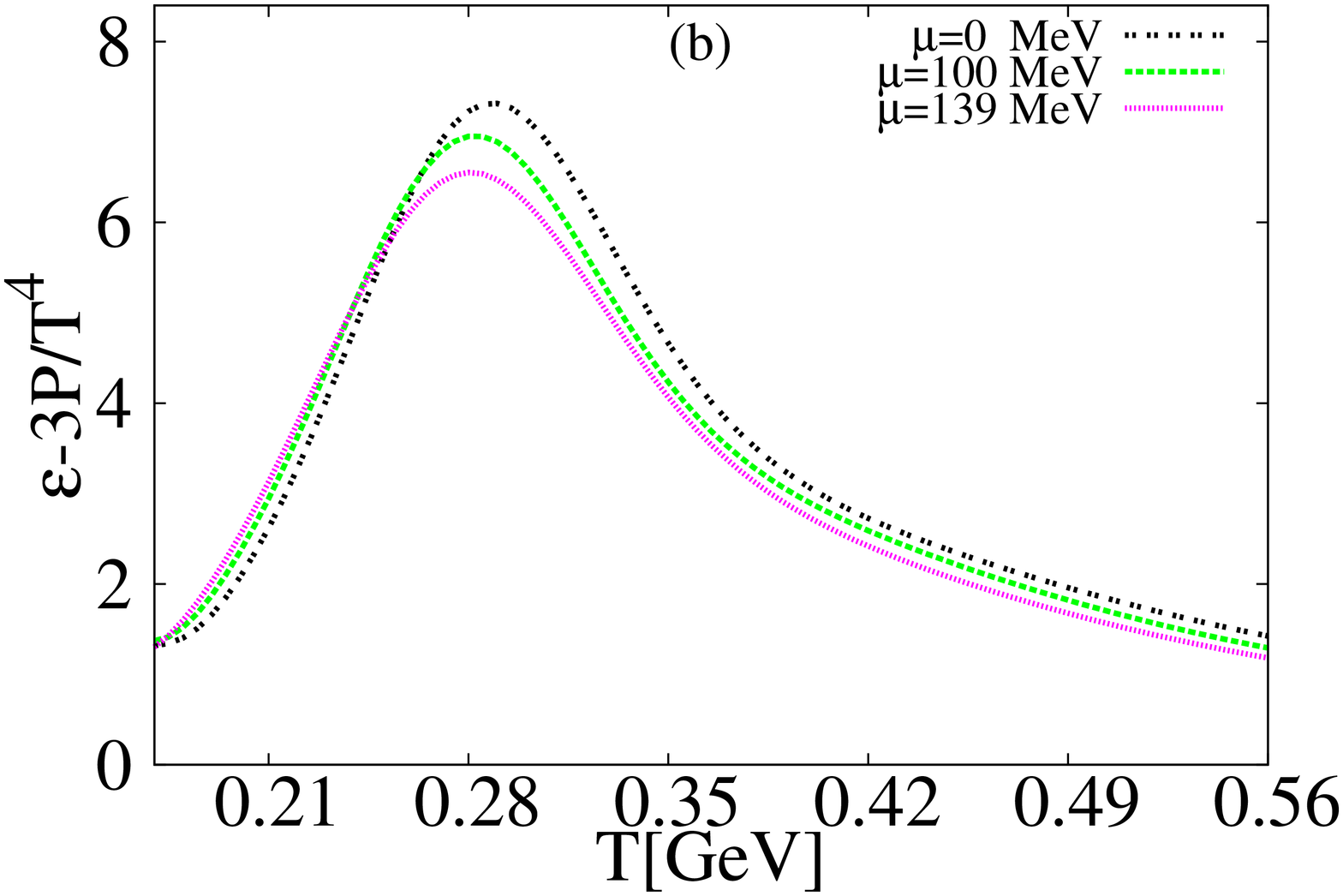}
\caption{(Color online) The left--hand panel shows the thermal behaviour of PLSM (curves) and lattice QCD trace anomaly (circles) \cite{HotQCD} (rectangles) \cite{QCDL} at vanishing chemical potential. The right--hand panel shows the thermal behaviour of PLSM trace anomaly but at different chemical potentials $\mu=\,0~$MeV (double--dotted curve), $100~$MeV (dashed curve) and $139~$MeV (dotted curve). \label{fig:Tr}}
}
\end{figure}

The trace anomaly of the energy--momentum tensor $ \mathcal{T}^{\nu \theta}$ also known as interaction measure reads 
\begin{eqnarray} \label{tr}
\Delta = \dfrac{ \mathcal{T}^{\nu \theta} }{T^4} = \dfrac{ \epsilon - 3 P }{T^4} =\,T \frac{\partial}{\partial T} \; \frac{P}{T^4}.
\end{eqnarray}
The results are given in Fig. \ref{fig:Tr}. Again, the general behavior is not totally absent, namely the values are small in hadronic phase and gradually increase in the region of phase transition (cross--over). They are decreasing in the deconfined phase. The trace anomaly shows a peak around the critical temperature $T_{\chi}$. This qualitatively agrees with the behavior observed in the lattice QCD calculations \cite{HotQCD,QCDL}. The agreement between PLSM and the lattice QCD \cite{HotQCD} is excellent \cite{Mao:2010}. Increasing Yukawa coupling comes up with a negligible improvement. Despite, the most recent lattice QCD calculations \cite{QCDL} are not reproducible by PLSM. 

In light of this, a short comparison between the two sets of lattice calculations is now in order.  Ref. \cite{QCDL} presented a full result for $2+1$ quark flavors, where all systematics are controlled, the quark masses are set to their physical values and the continuum extrapolation is carried out. Larger lattices and a Symanzik improved gauge and a stout--link improved staggered fermion action are implemented.  Depending on the exact  definition of the observables, the remnant of the chiral transition is at about $T_c=150$~MeV. Extending these results, the transition temperature  was also determined for small non-vanishing baryonic chemical potentials.

The lattice calculations in Ref. \cite{HotQCD} used $2+1$ quark flavors with physical strange quark mass and almost physical light quark--masses. The calculations have been performed with two different improved staggered fermion actions, the asqtad and p4 actions.  Overall, a good agreement between results obtained with these  two $O(a^2)$ improved staggered fermion discretization schemes is found.

At finite chemical potentials, the trace anomaly slightly increases (not shown here). At high $T$, the normalized trace anomaly gets close to the value of massless gas (at very high temperature, the Stefan--Boltzmann limit for  trance anomaly is zero).

\subsection{Comparing with Polyakov Nambu--Jeno--Lasino (PNLJ) model}
\label{subsec:thermo}
In this section, we introduce a comparison between the results obtained from PLSM and the Polyakov Nambu--Jeno--Lasino (PNLJ) model \cite{Fukushima:2003fw,Fukushima:2008wg} on the chiral condensates, the order parameters and the thermodynamic quantities. As introduced, some authors refer to LSM as Quark--Meson (QM) \cite{Fukushima:2003fw,Fukushima:2008wg}. We aim to find out how much the two models are close to each other and how they can reproduce  the lattice QCD calculations.

\begin{figure}[htb]
\centering{
\includegraphics[width=5.cm,angle=-90]{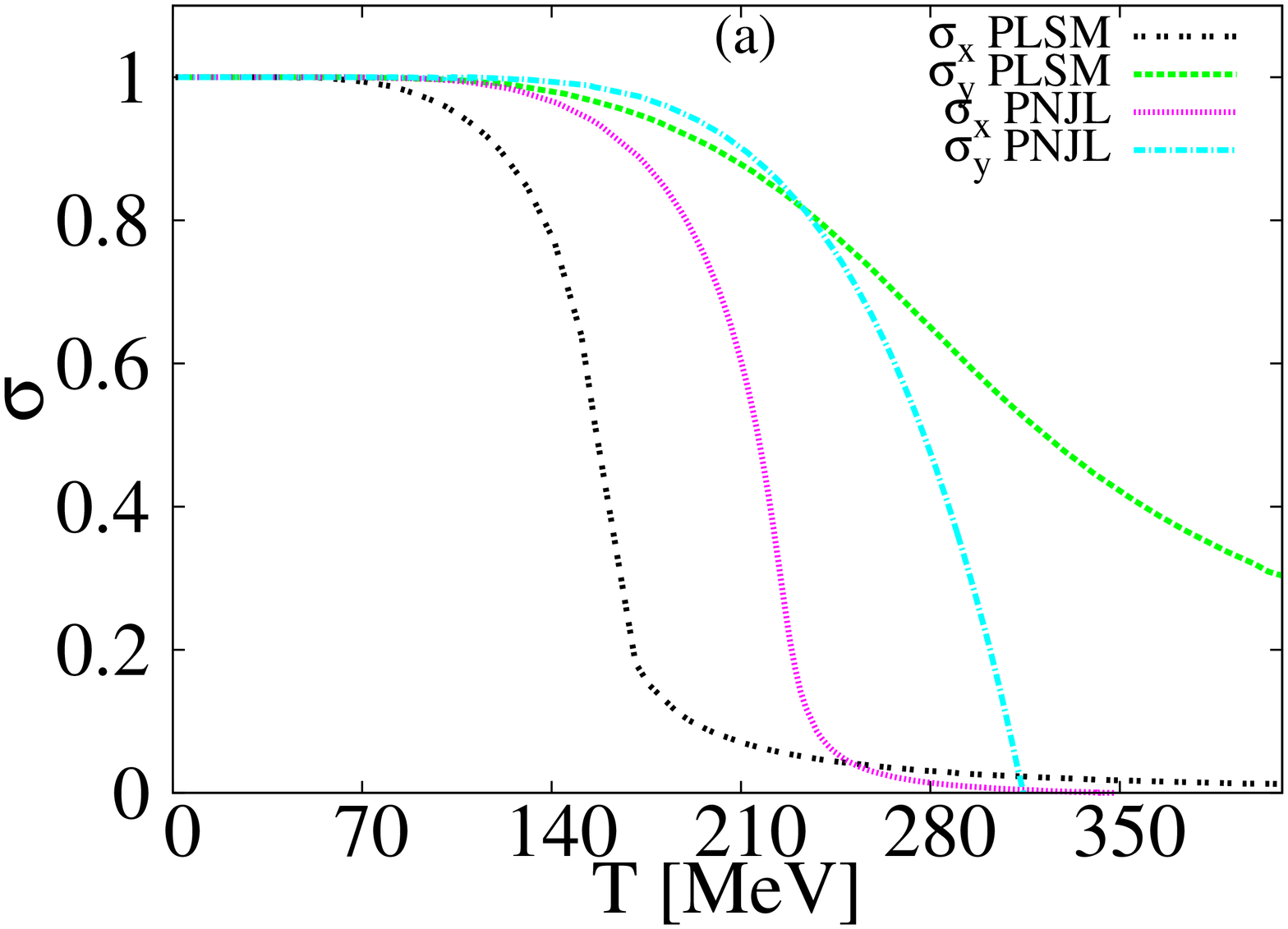}
\includegraphics[width=5.cm,angle=-90]{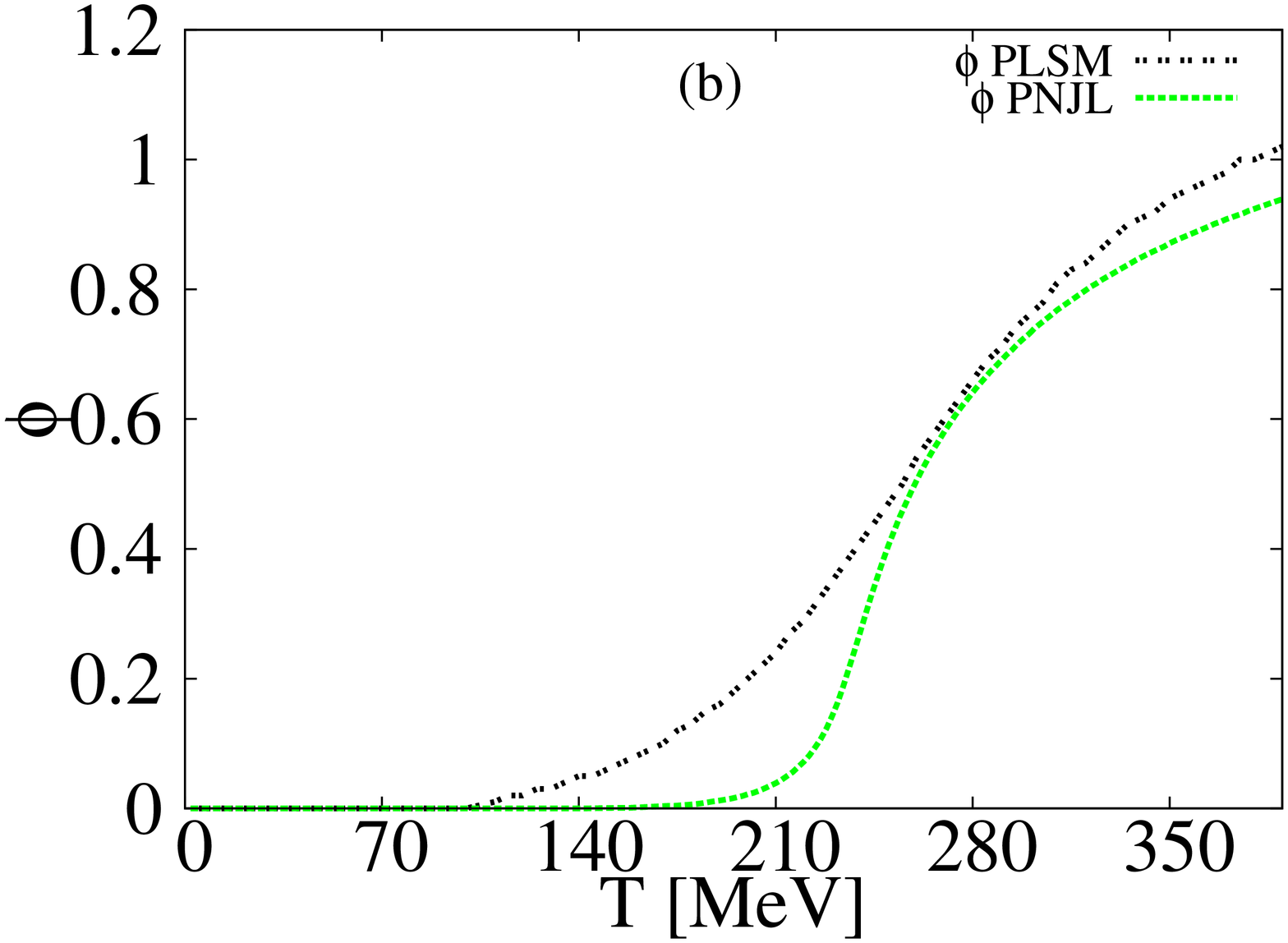}
\caption{(Color online) The left--hand panel shows the dependence of the normalized chiral condensates $\sigma_x $ and $\sigma_y$ in PLSM (double dashed and dotted curves, respectively) and $\sigma_x$ and $\sigma_y$ in PNJL (dashed and dotted dashed curves, respectively) on the temperature.  The right--hand panel shows the dependence of the order parameters of two models on $T$.  \label{fig:comp1} }
}
\end{figure}

\begin{figure}[htb]
\centering{
\includegraphics[width=5.cm,angle=-90]{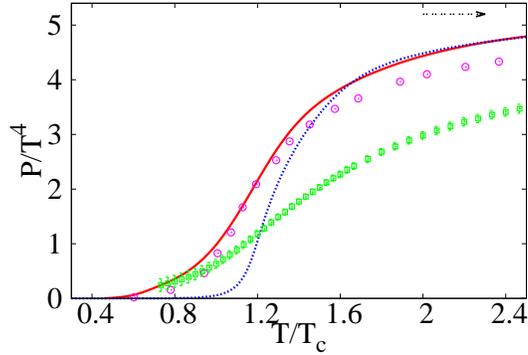}
\caption{(Color online) The thermal behavior of PLSM (solid curve), PNJL (dashed curve) and lattice QCD pressure (symbols) at vanishing chemical potential compared with the lattice QCD calculations (circles) \cite{HotQCD} (rectangles) \cite{QCDL}.\label{fig:comp2}}
}
\end{figure}

In left--hand panel of Fig. \ref{fig:comp1}, we compare between the thermal evolution of the normalized chiral condensates $\sigma_x$ and $\sigma_y$ in PLSM (double--dashed and dotted curves, respectively) and in PNJL (dashed and dash--dotted curves, respectively). We notice that the two models keep almost the same thermal behavior, especially in non--strange condensates.  With increasing $T$, PNJL show a saturation earlier that PLSM. Also, we find the PNJL condensates decrease faster than the PLSM condensates. This might be originated in thermodynamics of both models. Therefore, we show also another comparison for the thermodynamic pressure Fig. \ref{fig:comp2}.

The general behavior for the thermodynamic pressure exists in both models. Nevertheless, there are some differences, especially on how rapid is the jump between hadronic and partonic phase.  The PLSM fits well the lattice QCD calculations. Here, the jump seems to be slower than that of the PNJL. At high $T$, both models become coincident  with each other.

The PLSM and PNJL possess two order--parameters; one for strange and one for non--strange chiral condensates. Both are able to give hints about the chiral phase--transition. Because of the Polyakov loop potential, the two models possess also an additional order--parameter defining the deconfinement. At a fixed chemical potential, the thermal evolution of the strange and non--strange chiral condensates is used in estimating the critical temperature. It is apparent that this reflects whether a chiral phase--transition took place. At the same value of the chemical potential, we can also study the thermal evolution of the deconfinement order--parameter. Therefore, the thermal evolutions of these two quantities intersect with each other at a characterizing point representing the phase transition. This is the procedure we use to determine the critical temperature. The significances about this method are as follows. It determines a characteristic point for the chiral phase--transition and simultaneously distinguishes non--strange from strange chiral phase--transition.

In PNJL model,  there is a difference of about $5-10~$MeV in the critical temperature corresponding to chiral condensates and deconfinement Polyakov loop. We find such a difference is a little bit large \cite{TM}. This can occur because of the differences in the parameters. We have used large mass and the value of $T_0$ is small. This likely shifts Polyakov loop but in the opposite direction of the condensates. The parameters are listed in Ref. \cite{TM}.

\subsection{Higher Order Moments of Particle Multiplicity}
\label{sec:higher}

In this section, we introduce the first four non--normalized moments of particle multiplicity calculated in PLSM \cite{Schaefer:2011}. The thermal evolution is studied at different chemical potentials. Doing this, it turns to be possible to map out the chiral phase--diagram, for which we determine the irregular behavior in the higher moments as a function of $T$ and $\mu$. The possible fluctuations in the given quantity are related to non--monotonic behavior. Concretely, we use the fluctuations in the second order moment as an order parameter. In order words, at the peak of second order moment, the temperature and chemical potential shall be read out and then used to map out the chiral phase--transition. 

The higher order moments can be studied in different physical quantities, for example the particle multiplicity distribution \cite{Gupta}. Recently, the higher order moments of various multiplicity distributions have been reported by the STAR collaboration \cite{Xiaofeng,Tarnowsky} and the lattice QCD simulations~\cite{Schaefer:2011,Bielefeld}. An extensive study using statistical--thermal models has elaborated many features, especially that of the hadronic matter \cite{Tawfik:2013dba,Tawfik:2012si}.

There were many studies devoted to the higher order moments and the chiral phase transition of LSM and PLSM.  In Ref. \cite{Schaefer:2007d}, the thermodynamic quantises and the dimensionless first and second order moments were calculated. Also, it is conjectured that the chiral phase transition is related to the highest peak in the  dimensionless second order moment. The condensates were studied in dependence on temperature and chemical potential \cite{Schaefer:2008hk}.  Different Polyakov potentials have been assumed in evaluating the thermodynamic pressure, calculating its Taylor expansion and finally in mapping out the chiral phase transition \cite{Schaefer:2009ab}. Recently, these studies have been reviewed \cite{Schaefer:2011}. 

In section \ref{subsec:thermo}, different moments of the particle multiplicity are calculated. Non--normalized and normalized (with respect to $T$ or $\mu$) higher order moments are outlined in section \ref{subsec:nonnormalized} and \ref{subsec:normalized}, respectively. The chiral phase diagram is determined in section \ref{subsec:phasediagram} \cite{Tawfik:2013dba,Tawfik:2012si}.

\subsubsection{Non--normalized Higher Order Moments}
\label{subsec:nonnormalized}

\begin{figure}[htb]
\centering{
\includegraphics[width=5.cm,angle=-90]{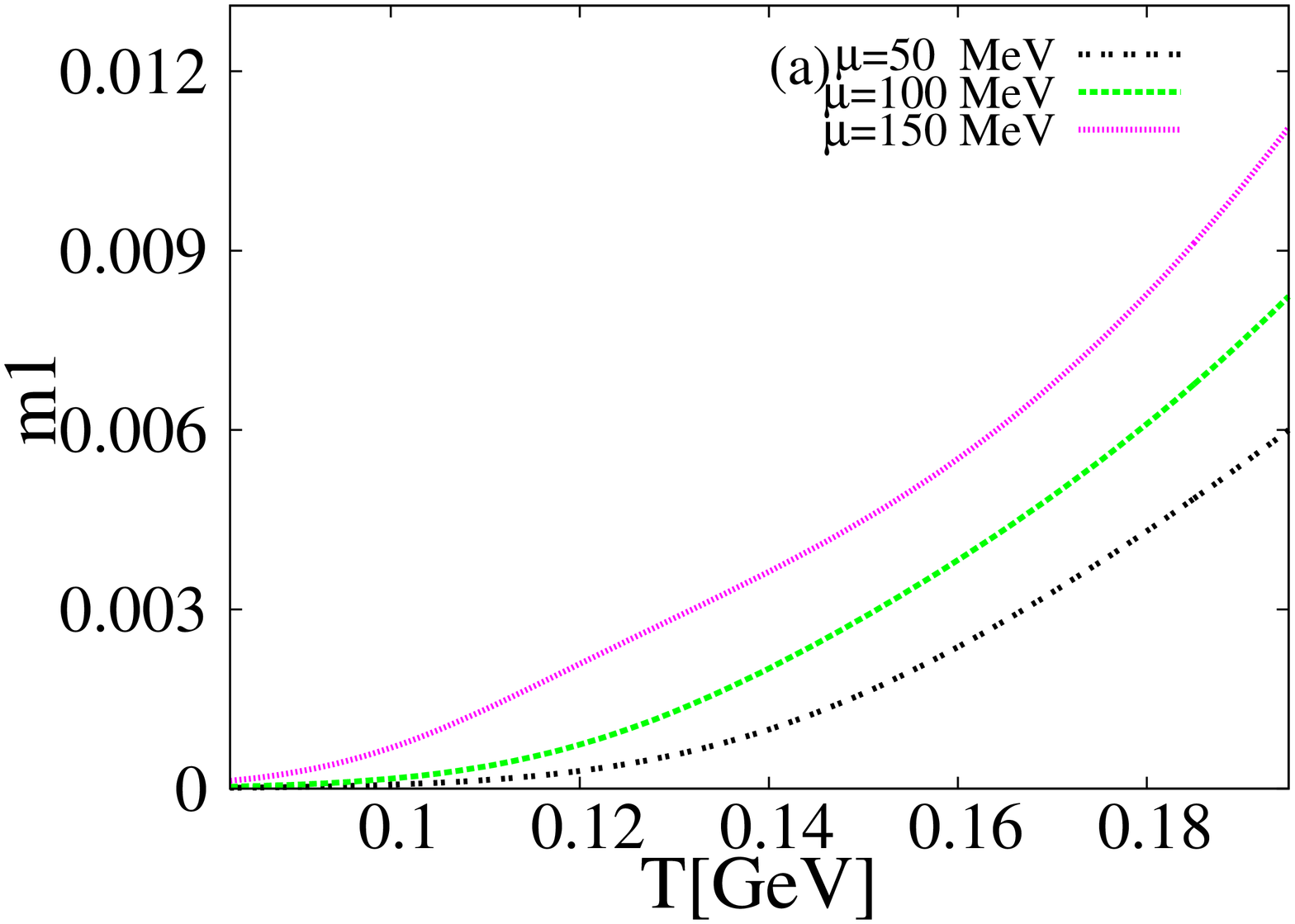}
\includegraphics[width=5.cm,angle=-90]{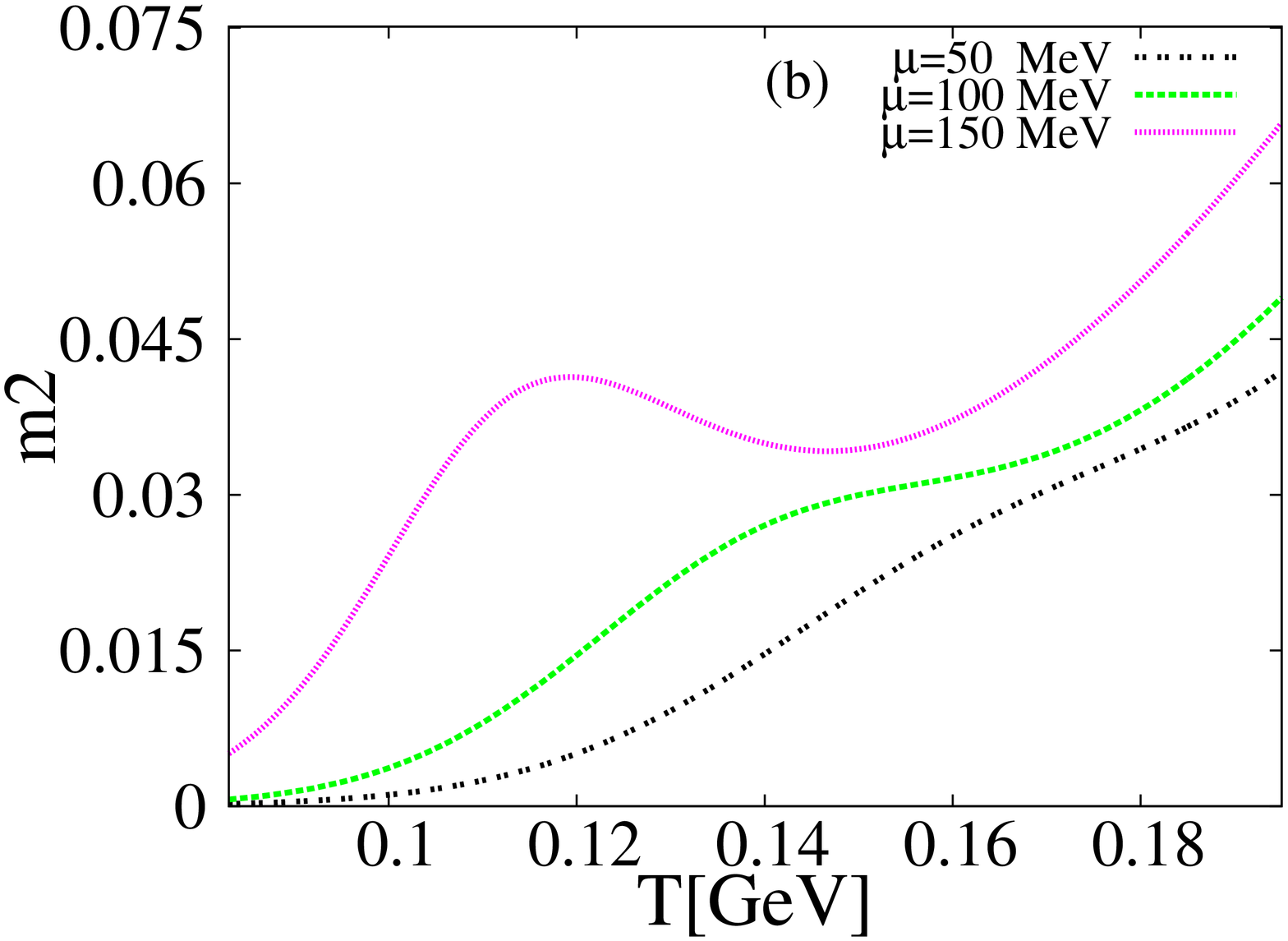}
\caption{(Color online) Left--hand panel: the first non--normalized moments are given as functions of temperature at $\mu=50~$MeV (double--dotted curve), $\mu=100~$MeV (dashed curve) and $\mu=150~$MeV (dotted curve). Right--hand panel: the same as in the left--hand panel but here for the second non--normalized moment.  \label{fig:m12} }
}
\end{figure}

\begin{figure}[htb]
\centering{
\includegraphics[width=5.cm,angle=-90]{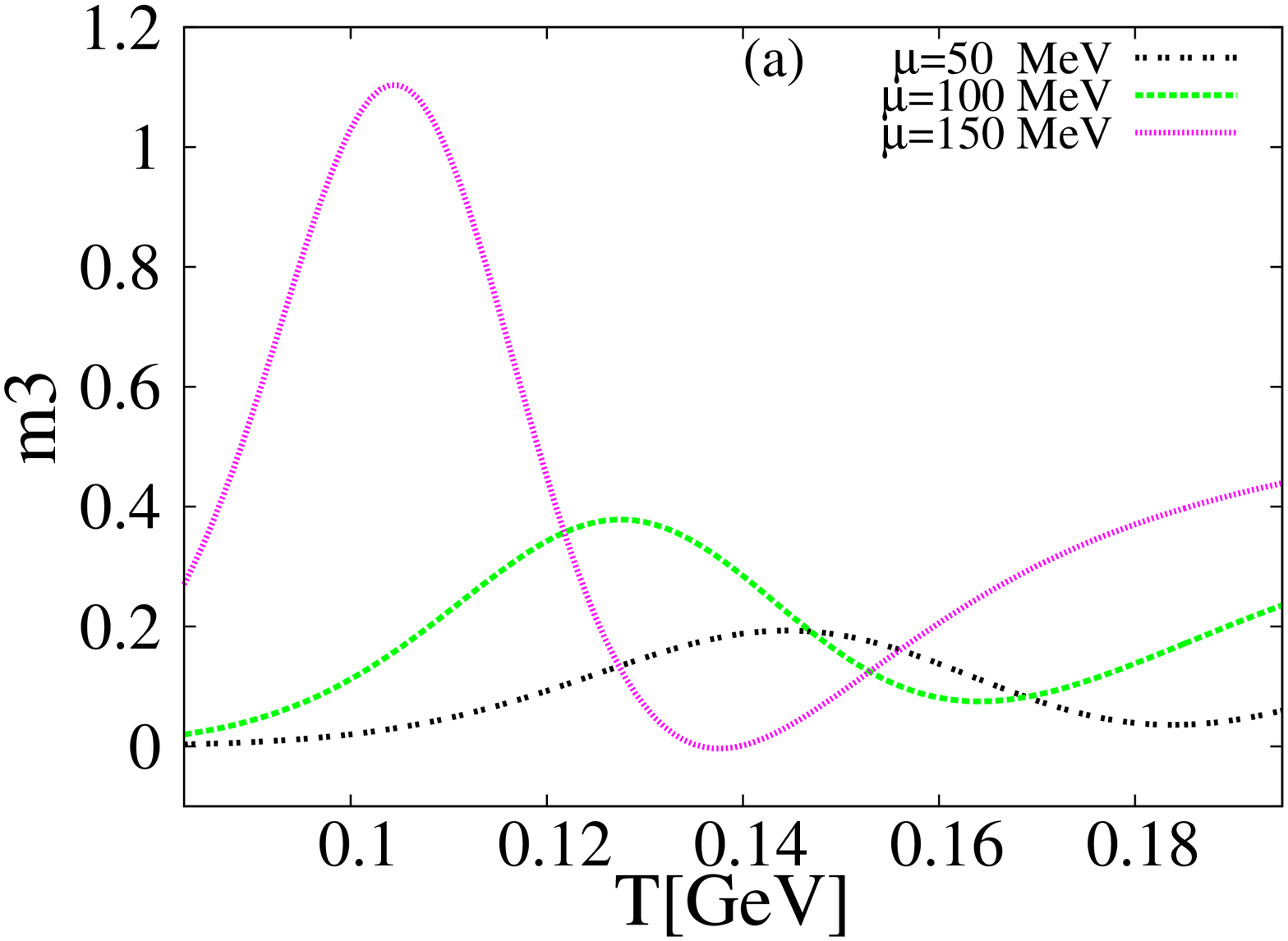}
\includegraphics[width=5.cm,angle=-90]{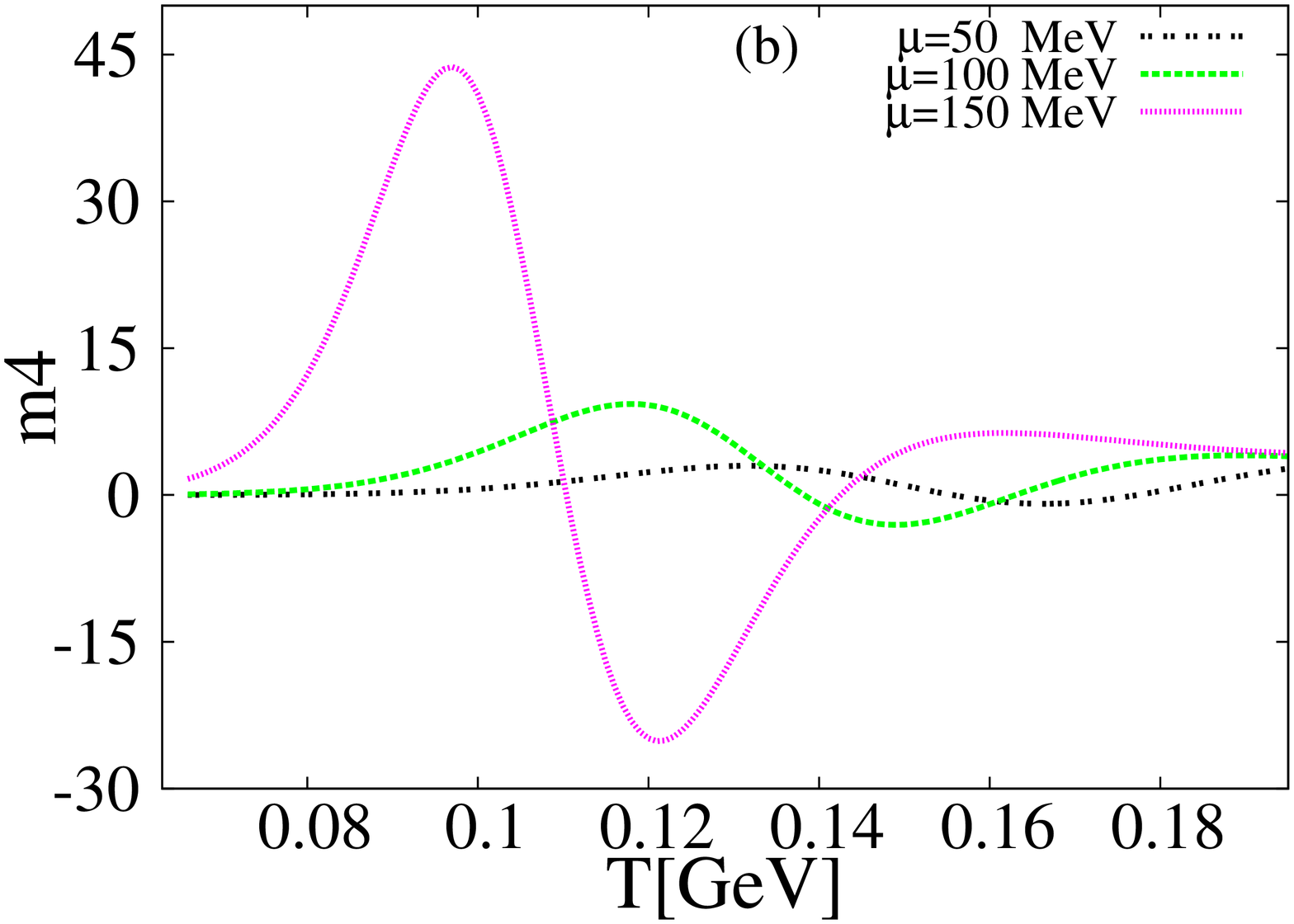}
\caption{(Color online) The same as in left--hand panel of Fig. \ref{fig:m12} but here for  third (left--hand panel) and fourth (right--hand panel)  non--normalized moments as functions of temperature at $\mu=50~$MeV (double-dotted curve), $\mu=100~$MeV (dashed curve) and $\mu=150~$MeV (dotted curve).   \label{fig:m34}  }
}
\end{figure}

The $i$-th higher--order moment of the particle multiplicity can be calculated from $\partial^i P/\partial \mu^i$, where $P$ is the pressure. An exact expression for $P$ is given in Eqs. (\ref{Pr}) and (\ref{newpotential}). All related expressions are also elaborated in sections \ref{sec:model} and \ref{sec:Results}. The first derivative describes the multiplicity distribution or the expectation operator, which can be utilized to estimate the number or multiplicity density, while the second order moment gives the variance of the given distribution. It is related to the susceptibility of the measurements. The first and second order moments are given in Append. \ref{appnd:1} and \ref{appnd:2}, respectively. The third order moment measures the lopsidedness of the distribution. Finally, the fourth order moment compares the tallness and skinny or shortness and squatness as shape of a certain measurement to its normal distribution. In the present work, the features of the first four order moments shall be characterized. Their dependence on $T$ and $\mu$ shall be analysed. 

\begin{figure}[htb]
\centering{
\includegraphics[width=5.cm,angle=-90]{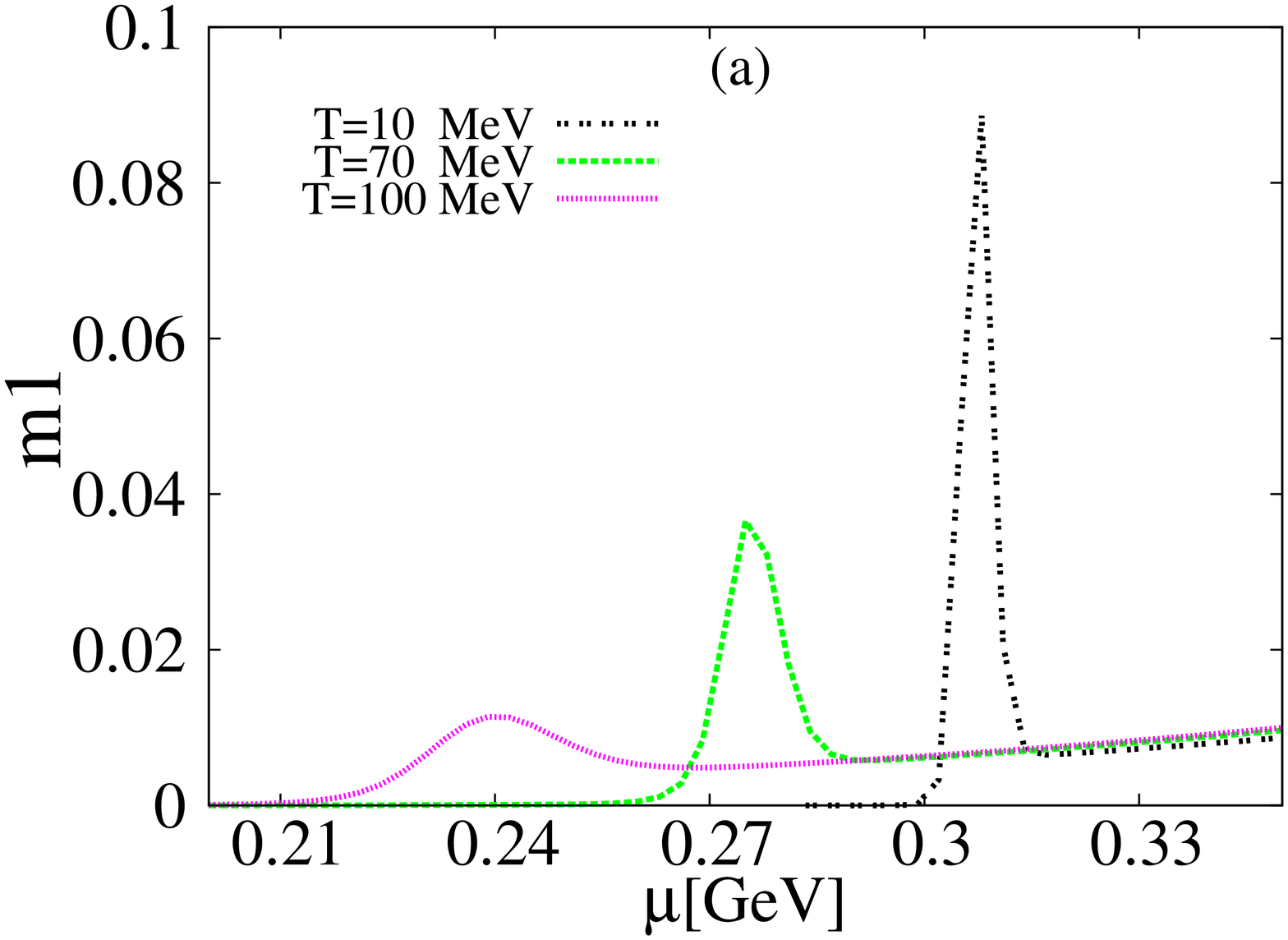}
\includegraphics[width=5.cm,angle=-90]{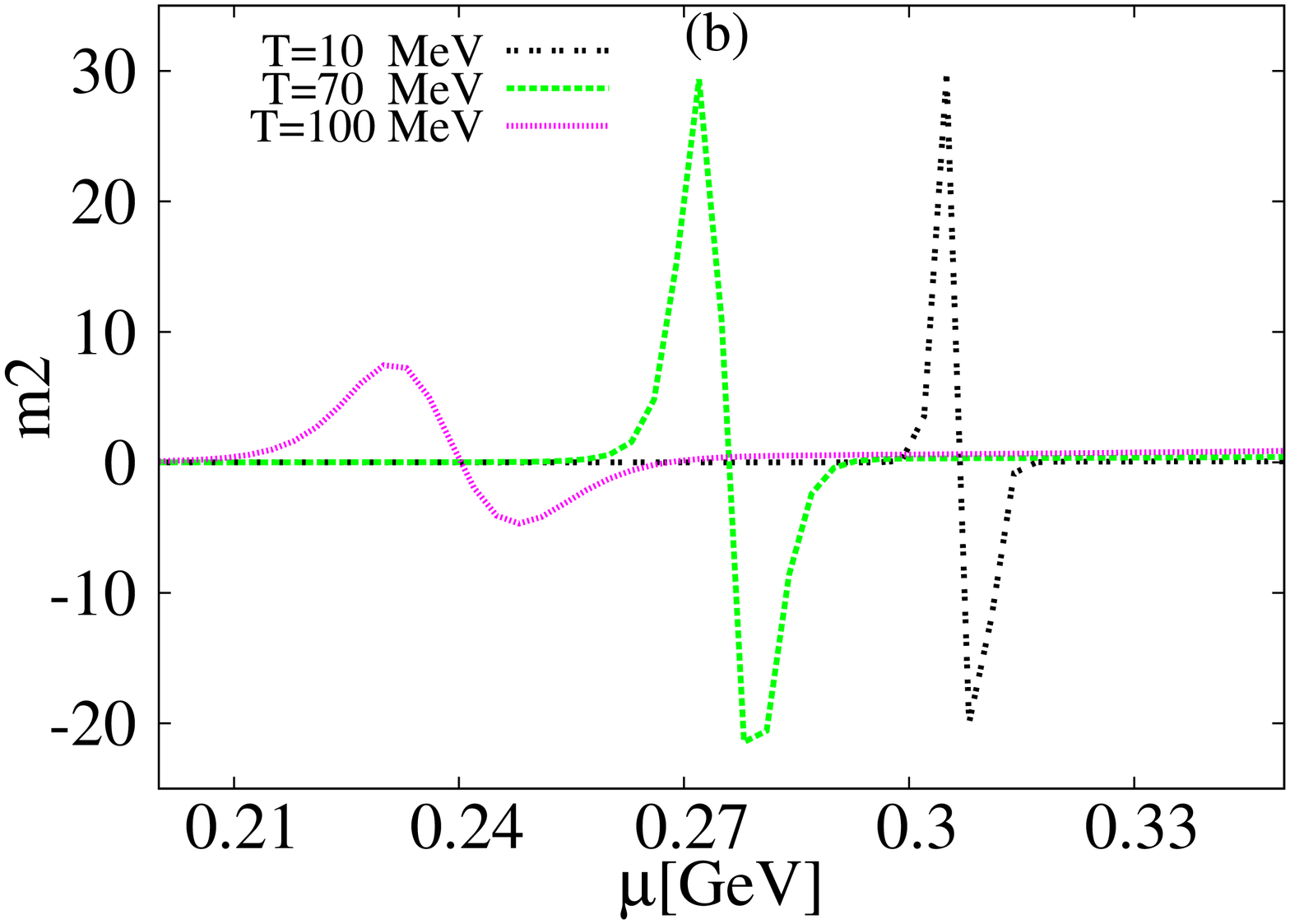}
\caption{(Color online) Left--hand panel: the non-normalized first moment is given as a function of the chemical potential at fixed temperatures $10~$MeV (double--dotted curve), $70~$MeV (dashed curve) and $100~$MeV (dotted curve). Right--hand panel: the same as in the left--hand panel second here for non--normalized second order moments.  \label{fig:m12nonwithMu}}
}
\end{figure}

\begin{figure}[htb]
\centering{
\includegraphics[width=5.cm,angle=-90]{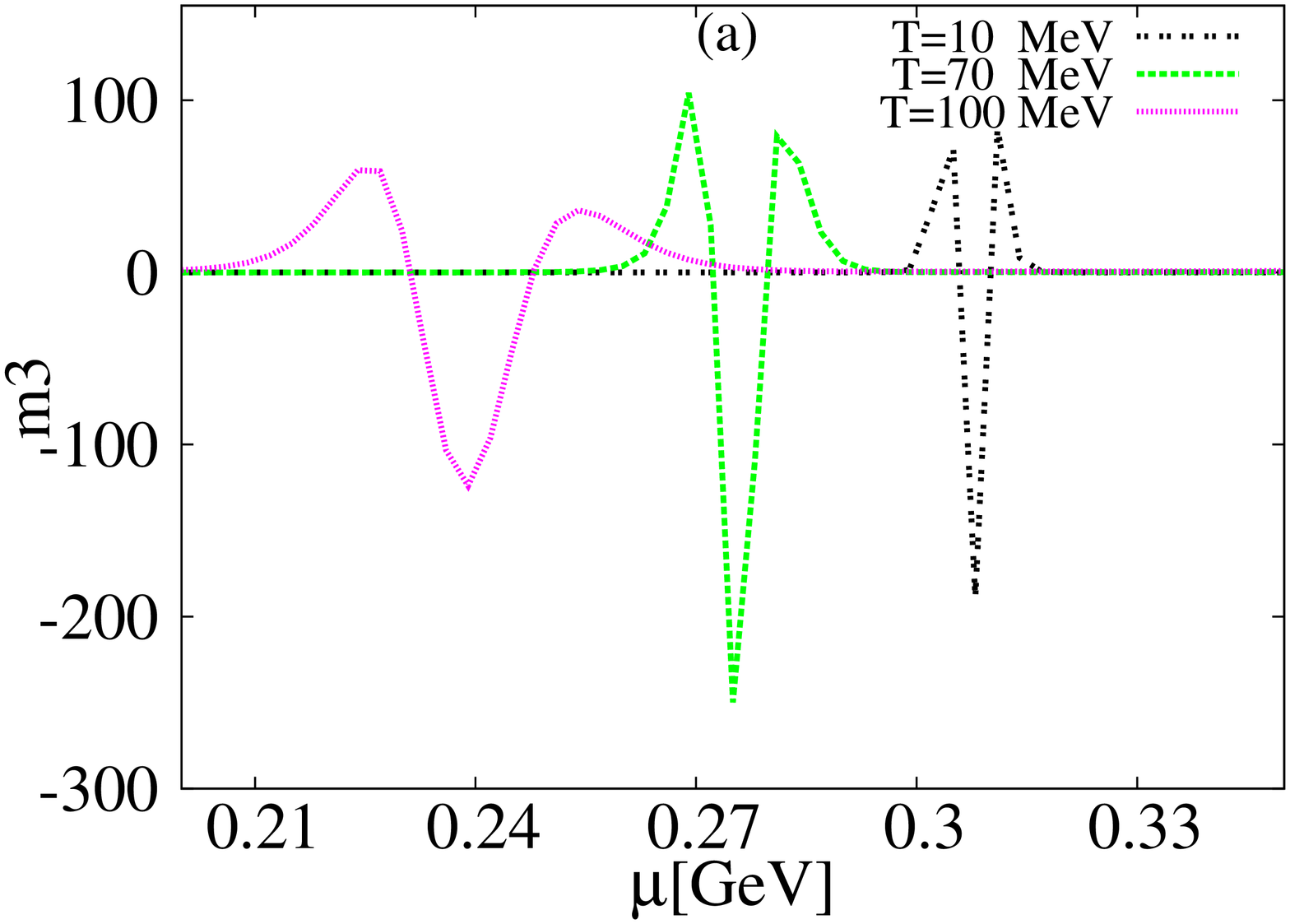}
\includegraphics[width=5.cm,angle=-90]{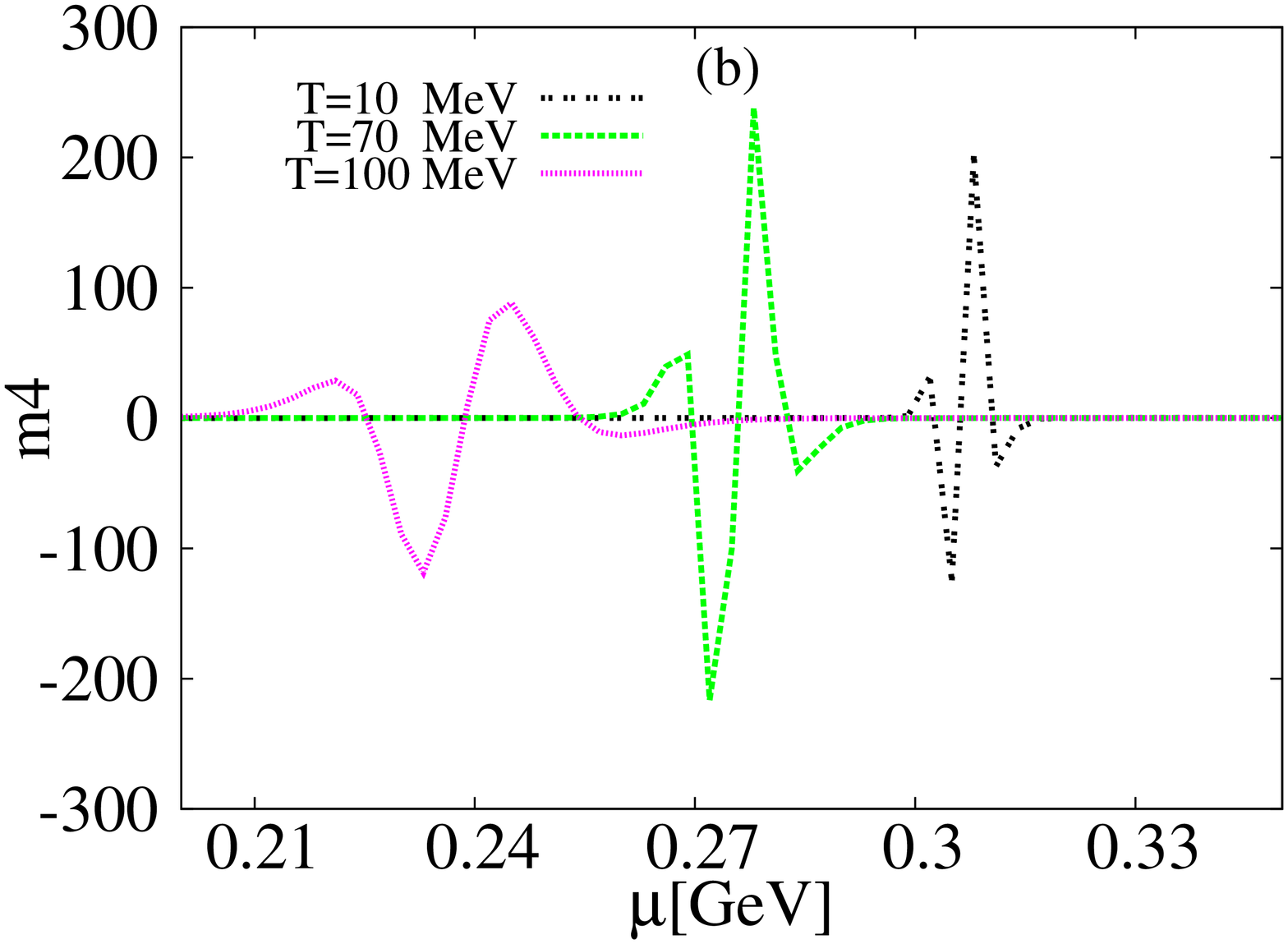}
\caption{(Color online) Left--hand panel: the third  non--normalized moment is given as a function of the chemical potential at fixed temperatures $10~$MeV (double--dotted curve here divided by $100$), $70~$MeV  (dashed curve here dided by $10$) and $100~$MeV (dotted curve). The right--hand panel gives the same as in the left--hand panel but for the fourth non--normalized moment. The double--dotted curve represent $T=10~$MeV  (divided by $100000$). The dashed and dotted curve give results at $T=70~$MeV (divided by $4000$) and $T=100~$MeV (divided by $200$), respectively. \label{fig:m34nonwithMu}}
}
\end{figure}

In Figs. \ref{fig:m12} and  \ref{fig:m34}, the first four non--normalized moments are summarized. They are given as functions of the temperature but as vanishing chemical potential. We find that increasing $T$ rapidly increases the four moments. Furthermore, the thermal dependence is obviously enhanced, when moving from lower to higher orders. Effects of the medium are also presented. Again, increasing the medium density (in terms of the chemical potential) leads to a further enhance in all moments. Comparing the four graphs with each others makes it clear that the enhancement is also related to the order, itself.  We also observe that the fluctuations are rapidly enhanced with increasing $T$, $\mu$ and the order. Figs. \ref{fig:m12nonwithMu} and \ref{fig:m34nonwithMu} reveal additional features.
 
To study the {\it pure} dependence on the medium density, we draw in Fig. \ref{fig:m12nonwithMu} the first (left--hand panel) and second (right-hand panel)  non--normalized moment as functions of chemical potential at fixed temperatures, $10~$MeV (double-dotted curve), $70~$MeV (dashed curve) and $100~$MeV (dotted curve).  Also, in Fig. \ref{fig:m34nonwithMu}, the third (left--hand panel) and fourth (right--hand panel)  non--normalized moments are given as functions of chemical potential at the same fixed temperatures. For a better appearance, some curves are scaled. We notice that increasing $T$ suppresses all moments, so that we should scale their curves. The peak is conjectured to reflect the critical transition. Accordingly, we find that increasing $T$ reduces the critical chemical potential. We find that the fluctuations become small with increasing $T$. 

At fixed chemical potentials, the thermal dependence of higher order moments become non--monotonic with increasing the orders and the chemical potentials, as well.  At fixed temperatures, the dependence on the medium density (related to the chemical potential) is also non--monotonic. So far, we conclude that PLSM seems to be sensitive to the fluctuations (peaks) accompanying different order moments. It is apparent that increasing $T$ weakens the fluctuations. As discussed in previous sections, the fluctuations of the higher order moments can be used to define the phase transitions. For instance, when taking into consideration the fluctuations in the first derivative of the order parameters, we can tackle the phase transitions for each type, either chiral or deconfinement, separately. Comparing the results from the fluctuations-- and intersections--method results in a small difference in the phase diagram.

\subsubsection{Dimensionless Higher Order Moments}
\label{subsec:normalized}
With normalized higher order moments, it is meant that the non--normalized moments are scaled to the standard deviation, $\sigma$ \cite{Tawfik:2012si}, where $\sigma$ is related to the susceptibility $\chi$ or the fluctuations. Due to sophisticated derivations and seeking for simplicity, we restrict the discussion to dimensionless higher order moments. This can be done with respect to the temperature, $T$, or the chemical potential, $\mu$. It is conjectured that the phase transition could also be indicated by large fluctuations in the dimensionless moments. This can be implemented in order to map out the chiral phase--transition. Practically, this step simply refers to the dimension, in which the phase diagram should be analysed. The dependence of the higher moments on $T$ is practically related to a scan in the $T$ dimension. Also, their dependence on $\mu$ is practically related to a second scan in the $\mu$ dimension.

\begin{figure}[htb]
\centering{
\includegraphics[width=5.cm,angle=-90]{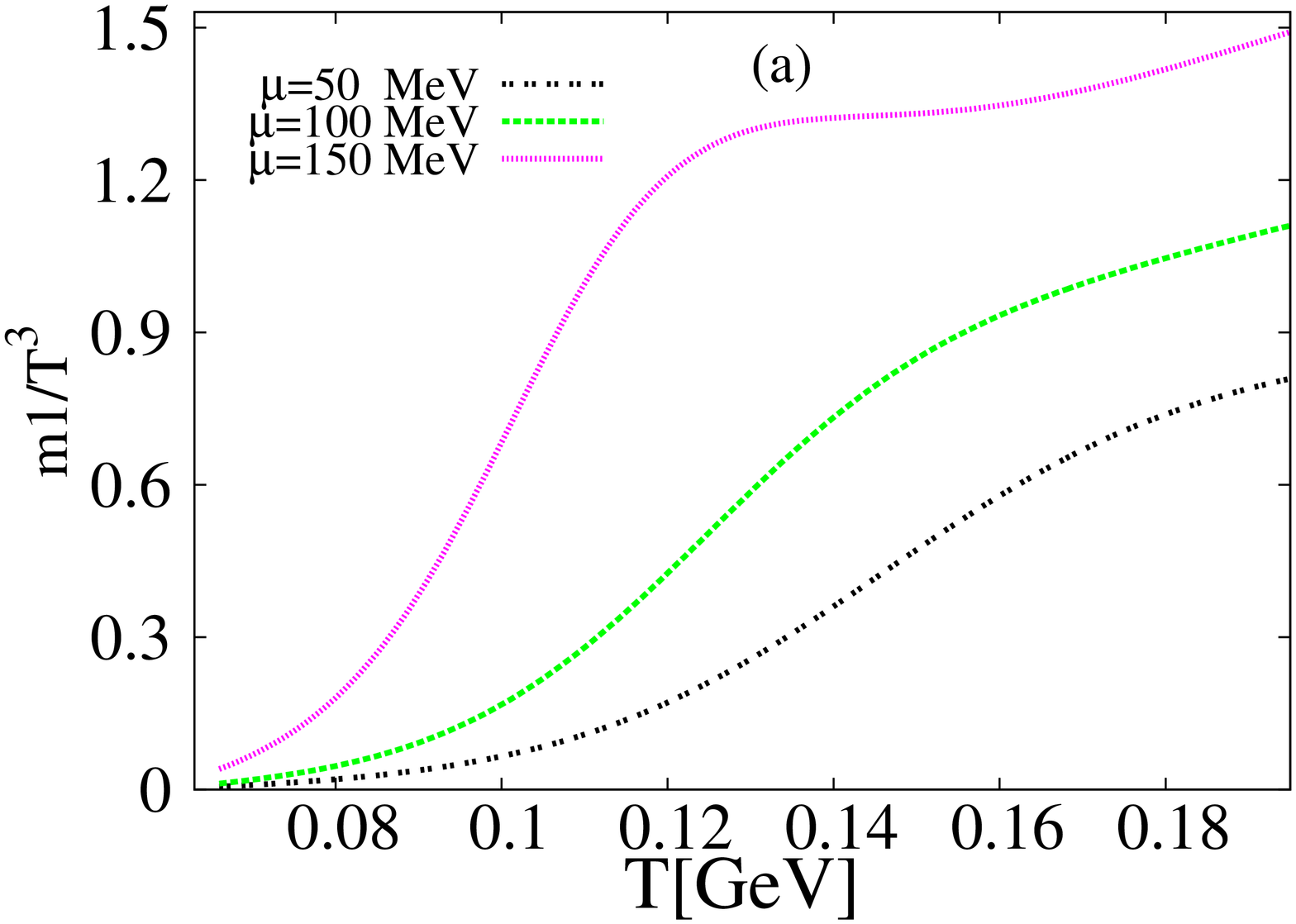}
\includegraphics[width=5.cm,angle=-90]{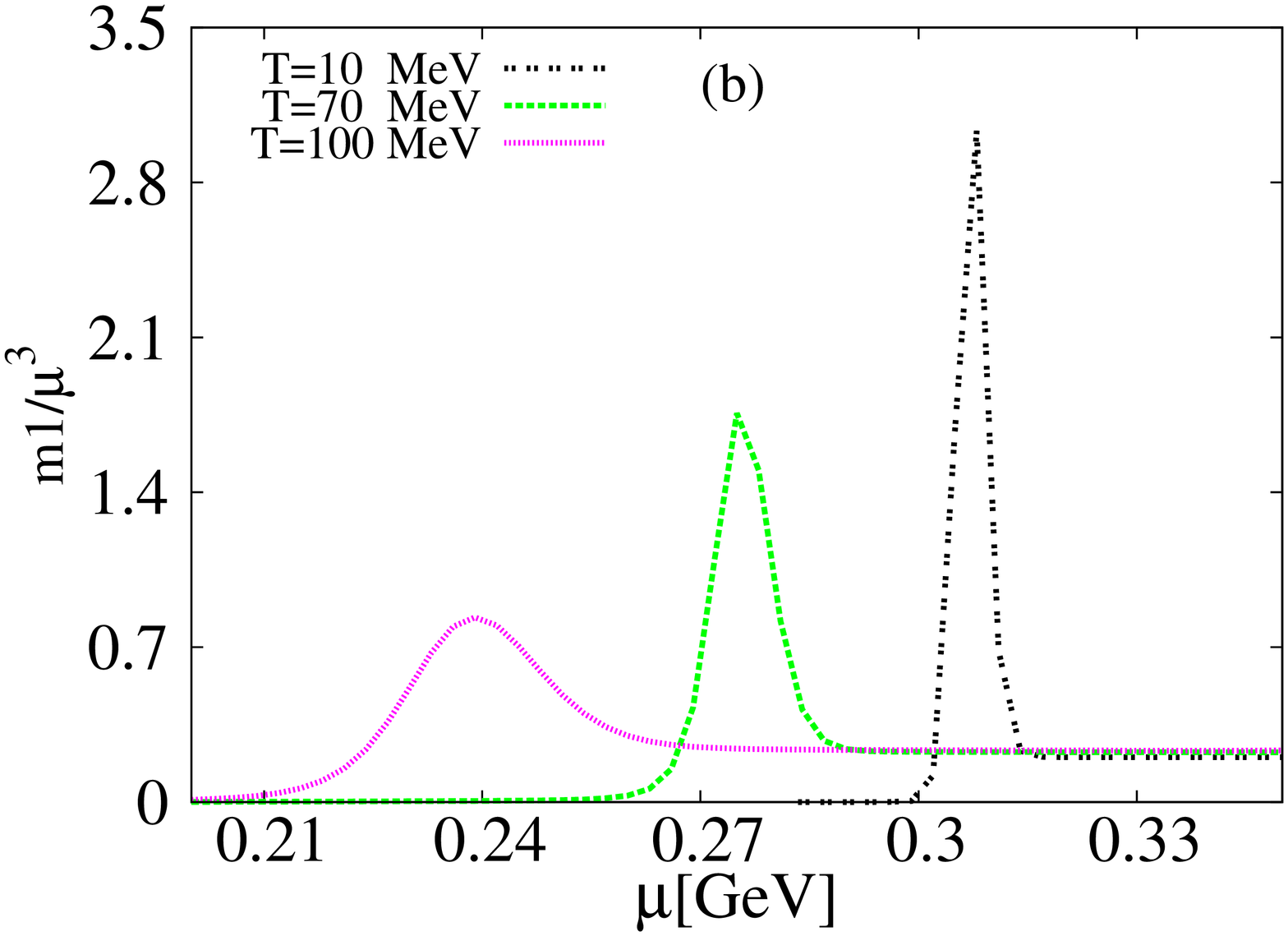}
\caption{(Color online) Left--hand panel: $n/T^3$ is given in dependence on  temperature at fixed chemical potentials $\mu = 5~$MeV (double--dotted curve), $\mu=100~$MeV (dashed curve) and $\mu=150~$MeV (dotted curve). The right--hand panel gives $n/\mu^3$ as a function of chemical potential at fixed temperatures $T=10~$MeV (double--dotted curve), $T=70~$MeV (dashed curve) and $T=100~$MeV (dotted curve). \label{fig:m1normwithTMu}}
}
\end{figure}

As given in Appendix \ref{appnd:1}, the first order moment, $n$, can be normalized to $T^3$ or to $\mu^3$.  In the left--hand panel of Fig. \ref{fig:m1normwithTMu}, $n/T^3$ is given as a function of $T$ at fixed chemical potentials $\mu=50$, $100$ and $150~$MeV. The right--hand panel shows $n/\mu^3$ as a function of $\mu$ but at fixed temperatures $T=10$, $70$ and $100~$MeV. It is obvious that increasing $\mu$ increases the value of $n/T^3$, compare with the left--hand panel. The same dependence was found in the lattice QCD calculations  \cite{HotQCD,Karsch2009a,QCDL}. As noticed in Fig. \ref{fig:m12nonwithMu}, the dependence on $\mu$ seems to unveil essential fluctuations. Contrary to the thermal evolution of $n$ or $n/T^3$, the dependence of $n/T^3$ on $\mu$ is accompanied with drastic fluctuations (large peaks). The latter gets larger with increasing the orders and/or the temperatures. The dependence of $n/\mu^3$ on $\mu$ is presented in the right-hand panel of Fig. \ref{fig:m1normwithTMu}. Comparing this with the left--hand panel of Fig. \ref{fig:m12nonwithMu} makes it clear that the fluctuations in the earlier case get enhanced. Also, we notice that increasing $T$ reduces the height of the peak  (fluctuation).  The dependence of $n/T^3$ on $T$ is practically related to scanning the phase diagram in $T$ dimension. Also, the dependence of $n/\mu^3$ on $\mu$ is practically related to scanning the phase diagram in the $\mu$ dimension. Also, we observe that increasing $T$ obviously decreases the value of $n/\mu^3$ and weakens the sharpness of the peaks.

\begin{figure}[htb]
\centering{
\includegraphics[width=5.cm,angle=-90]{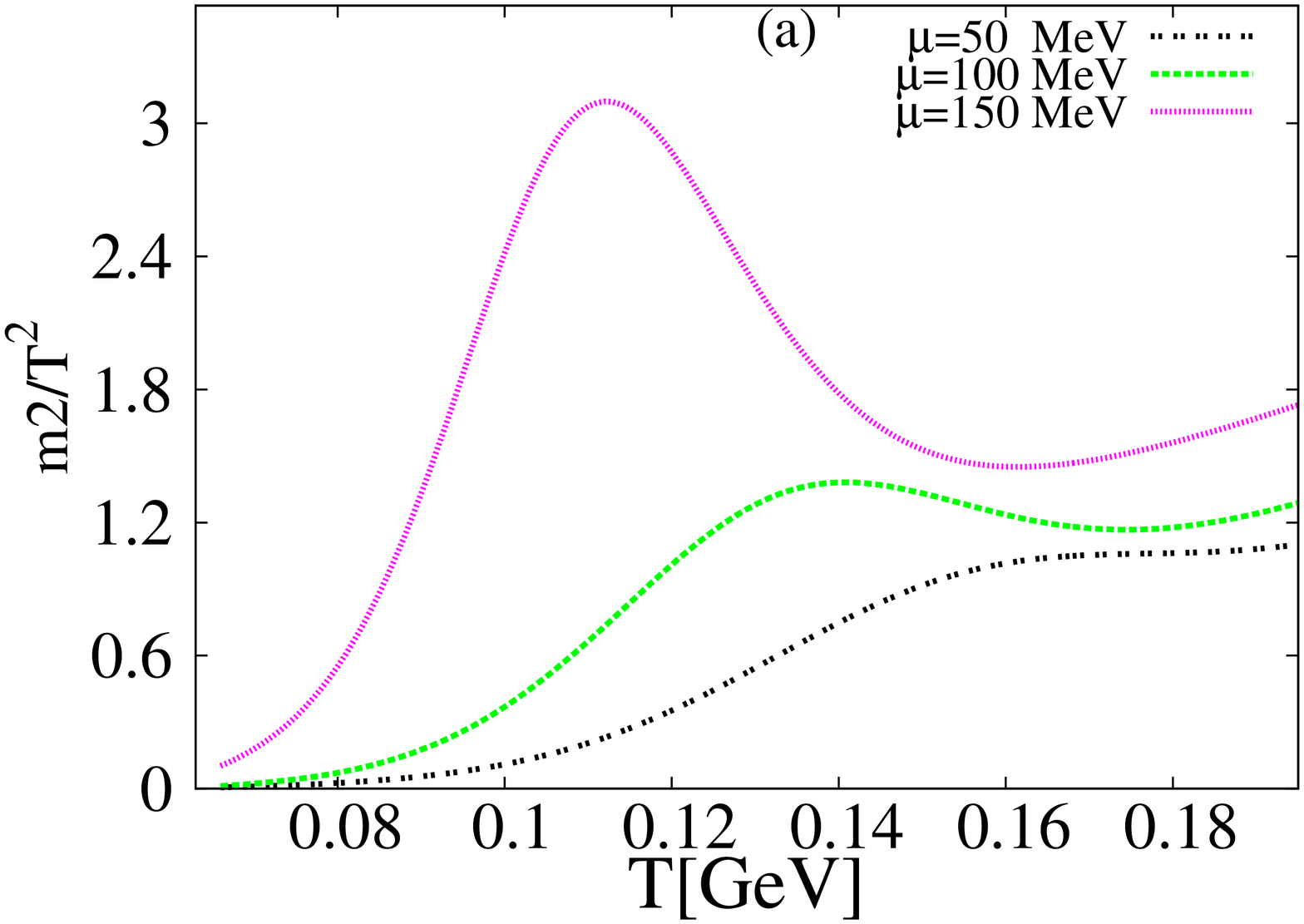}
\includegraphics[width=5.cm,angle=-90]{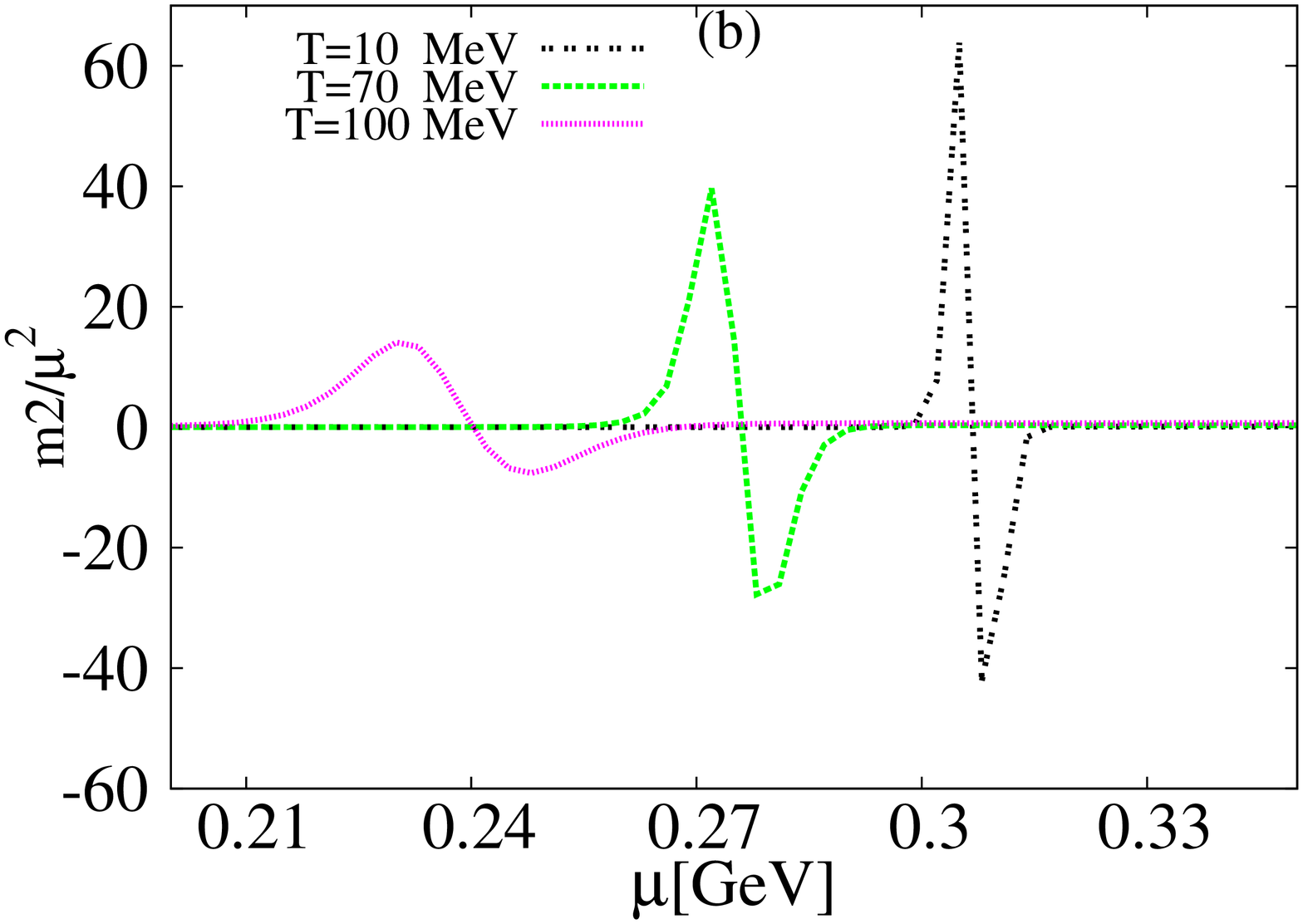}
\caption{(Color online) Left--hand panel: the thermal evolution of the second order moments scaled to the temperature is given at different chemical potentials $\mu=50~$MeV (double-dotted curve), $\mu=100~$MeV (dashed curve) and $\mu=150~$MeV (dotted curve).  Right--hand panel: the density dependence of $m_2/\mu^2$ at fixed temperatures, $10~$MeV (double--dotted curve here divided by $5$), $70~$MeV (dashed curve here divided by $2$) and $100~$MeV (dotted curve).  \label{fig:m23normwithT}}
}
\end{figure}

As given in Appendix \ref{appnd:2}, the second order moment can be scaled to $T^2$ or $\mu^2$. Fig. \ref{fig:m23normwithT} shows the thermal dependence of $m_2/T^2$ (left-hand panel) and medium dependence of $m_2/\mu^2$ (right--hand panel). Increasing $\mu$ increases the value of $m_2/T^2$, while Increasing $T$ decreases the value of $m_2/\mu^2$. Comparing this with the thermal evolution of $n/T^3$ (left--hand panel of Fig. \ref{fig:m1normwithTMu}) leads to the conclusion that higher order moments should be accompanied with increasing fluctuations.  Increasing $T$ obviously decreases the value of $m_2/\mu^2$ and weakens the sharpness of the peaks. 

\begin{figure}[htb]
\centering{
\includegraphics[width=5.cm,angle=-90]{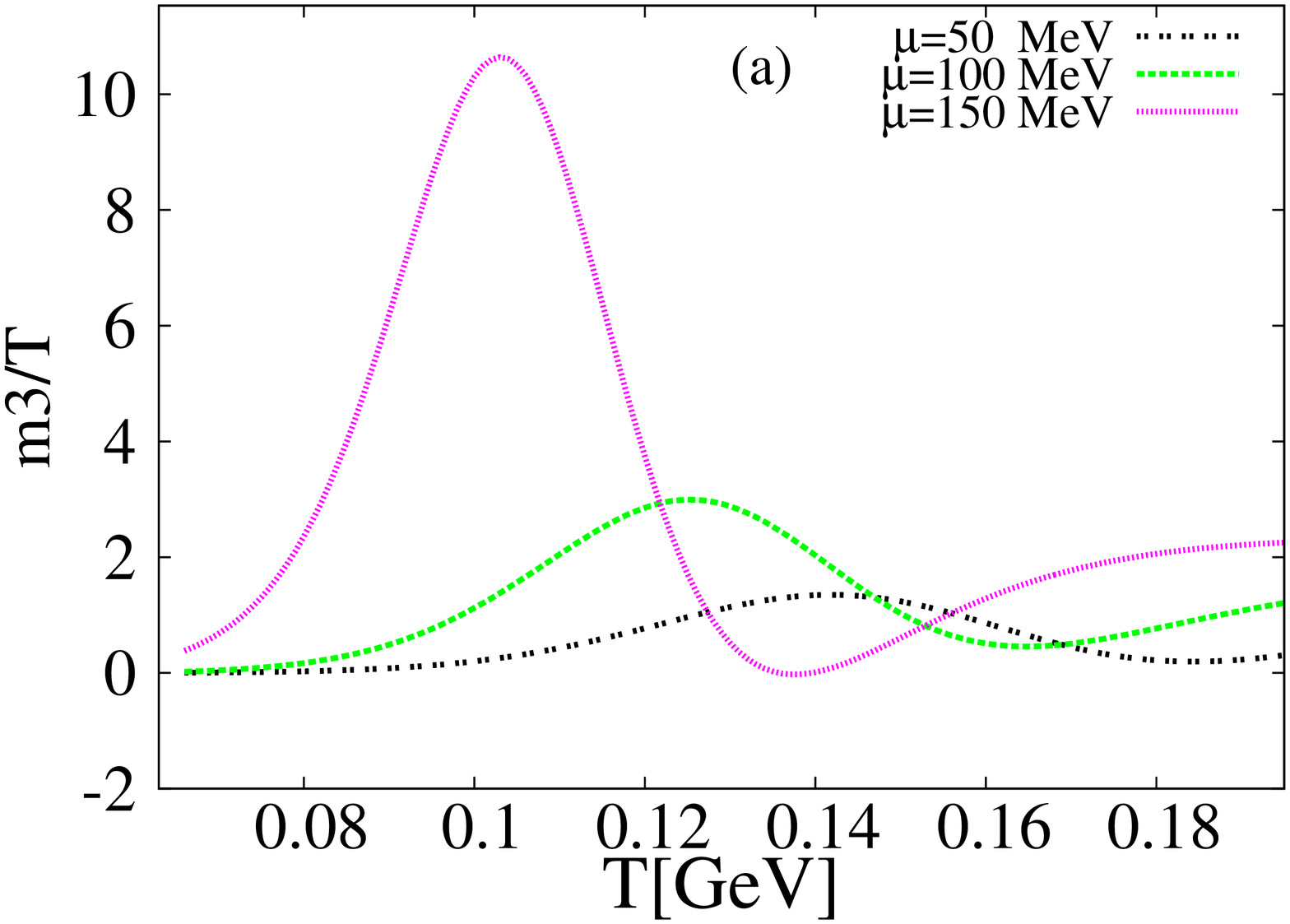}
\includegraphics[width=5.cm,angle=-90]{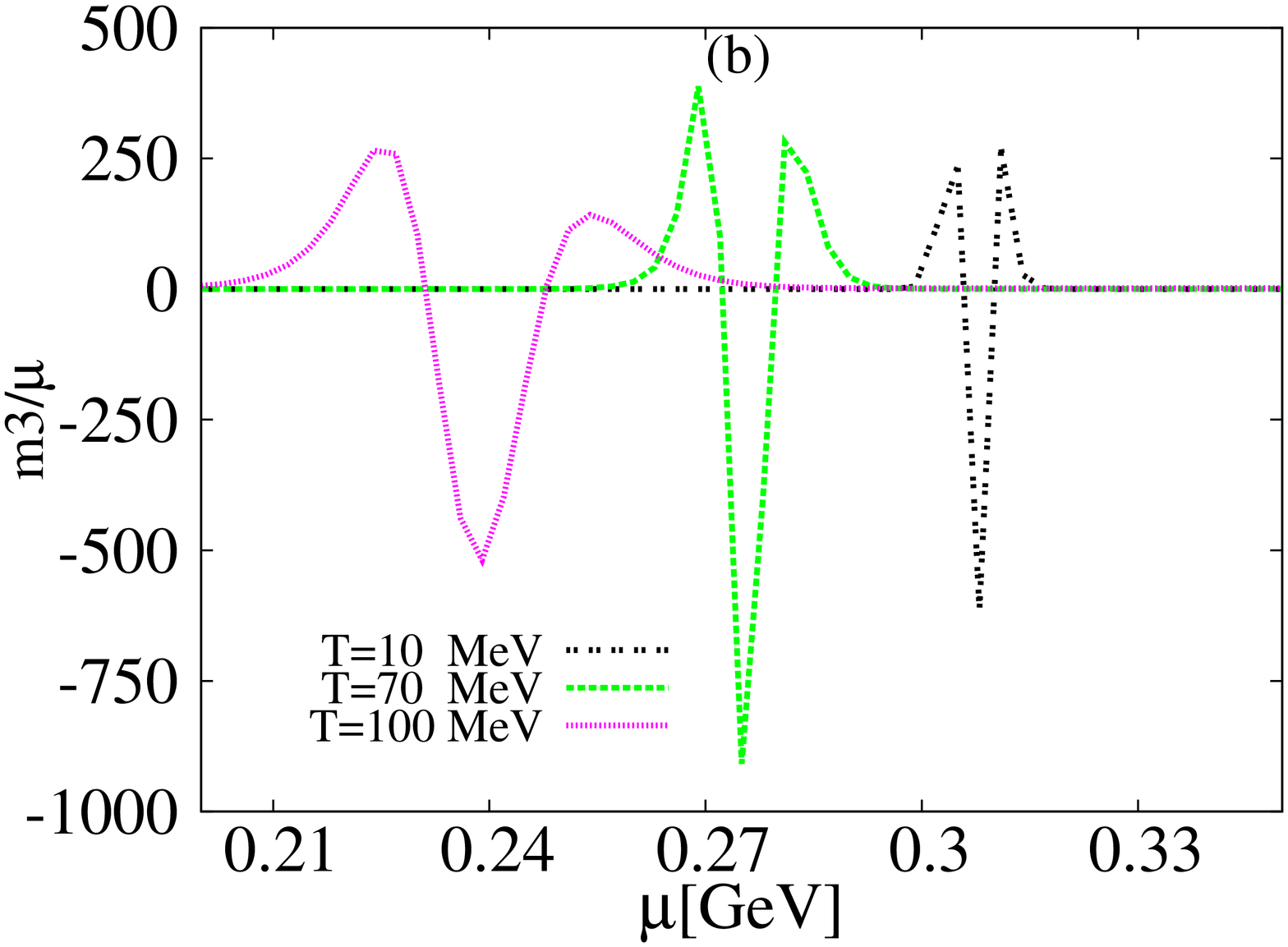}
\caption{(Color online) Left--hand panel: the thermal evolution of the third order moment scaled to temperature is evaluated at different chemical potentials $\mu=50~$MeV (double--dotted curve), $\mu=100~$MeV (dashed curve) and $\mu=150~$MeV (dotted curve). Right--hand panel: the medium dependence of the third order moment scaled to the chemical potential is given at fixed temperatures $10~$MeV (double--dotted curve here divided by $100$), $70~$MeV  (dashed curve here divided by $10$) and $100~$MeV (dotted curve).   \label{fig:m3normwithMu}}
}
\end{figure}

In the left--hand panel of Fig. \ref{fig:m3normwithMu}, the thermal evolution of the third order moment scaled to temperature is given at different chemical potentials $\mu=50~$MeV (double--dotted curve), $\mu=100~$MeV (dashed curve) and $\mu=150~$MeV (dotted curve). Increasing $\mu$ is accompanied by a remarkable increase in $m_3/T$ and enhanced fluctuations. Comparing with the previous two figures, it is apparent that the fluctuations increases with raising the order of the moment. The right--hand panel shows the medium dependence of the third order moment scaled to the chemical potential at fixed temperatures $10~$MeV (double-dotted curve here divided by $100$), $70~$MeV (dashed curve here divided by $10$) and $100~$MeV (dotted curve). Again, increasing $T$ reduces the height of the peak 
(this was observed in the previous two figures) and weakens the sharpness of the peaks. 

The thermal and dense dependence of the fourth order moment was given in Figs. \ref{fig:m34} (right--hand panel) and \ref{fig:m34nonwithMu} (right panel), respectively. Additional to it and as illustrated in Figs. \ref{fig:m1normwithTMu}-\ref{fig:m3normwithMu}, the systematic dependence of the first four order moments on temperature and chemical potential is confirmed.

\subsection{Chiral  Phase Transition and Freeze-out Diagram}
\label{subsec:phasediagram}

As introduced in section \ref{sec:higher},  the higher order moments are sensitive to the fluctuations in $T$ and $\mu$ dimensions. This is valid for non--normalized moments as well as for the ones normalized to $T$ or $\mu$. The fluctuations likely appear in case of a drastic change in the degrees of freedom, symmetry change/restoration or in the dynamics deriving the system out of equilibrium \cite{critPT}. In the present work, we utilize possible fluctuations accompanying the normalized second order moment \cite{Schaefer:2008hk,Karsch2009a} in mapping out the chiral phase--transition. The procedure of determining {\it quasi} critical temperature from the second moment has been discussed  \cite{Karsch2009a}.  Accordingly, we observe that the peak in each figure is characterized by $T$- and $\mu$-values, where chiral phase transition is conjectured to take place. In the left-hand panel of Fig. \ref{fig:m23normwithT}, the dependence of $\chi/\mu^2$ on the chemical potential at fixed temperatures is depicted. We found that $\mu$-values of the peaks vary with $T$. We scan this dependence at different values of $T$. Then, we follow the scheme to determine $T$ and $\mu$, at which $\chi/\mu^2$ gets maximum.

\begin{figure}[htb]
\centering{
\includegraphics[width=6.cm,angle=-90]{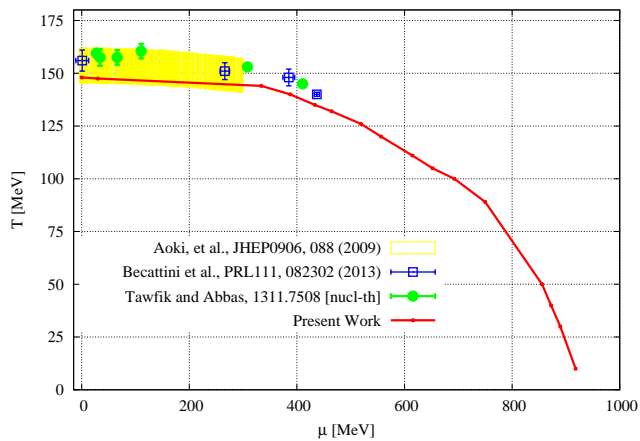}
\caption{(Color online) The PLSM $T$-$\mu$ chiral phase diagram (lines with points), with which the freeze-out parameters deduced from lattice QCD calculations \cite{KarschA,LQCDA} (band) and that from different thermal models \cite{Tawfik:2013bza,SHMUrQM} (symbold) are compared. 
\label{fig:TandMub} }
}
\end{figure}

In Fig. \ref{fig:TandMub}, the relation between $T$ and $\mu$, at which the second order moment gets maximum value is illustrated. We compare this with the  freeze-out parameters, the temperature and the baryon chemical potential. Also, the lattice QCD results are compared with. 

We observe that the PLSM chiral boundary is apparently positioned in the lower band of the lattice QCD calculations  \cite{KarschA,LQCDA}. The freeze-out parameters deduced from the lattice QCD calculations  \cite{KarschA,LQCDA} (band) and that from different thermal models \cite{SHMUrQM} (open symbols) \cite{Tawfik:2013bza} (solid symbols) are slightly large. The freeze-out results at large $\mu$ are currently under debate \cite{Tawfik:2013bza}. Therefore, it is too early to make conclusion about the whole chiral phase diagram. It is apparent, that the freeze-out boundary and chiral phase transition are almost non-distinguishable, especially at small $\mu$.

The chiral phase--diagram can be determined by different methods. For PLSM and PNJL, as they possess two order parameters; one for strange and one for non--strange chiral condensates. This order parameter gives hints about the chiral phase--transition. Furthermore, PLSM and PNJL possess deconfinement order parameter because of the Polyakov loop potential. From strange and non--strange chiral condensates which  can be deduced as functions of temperatures at fixed chemical potential. At the same value of chemical potential, we can also deduce the deconfinement order parameter as a function of temperatures.Thus, the thermal evolution of these two quantities intersect with each other at a characterizing point representing the phase transition. This method is sensitive to the chiral phase--transition.

Another method based on the fluctuations of some derivative quantities  can determine other types of the phase transitions. For instance, the first derivative of the chiral condensate and the Polyakov loop with respect to the  temperature  fluctuates with changing chemical potential. When this step is repeated with different temperatures, the chiral phase--diagram can be drawn. A third method based on fluctuations of the higher order moments is also implemented in the present work.

\section{Conclusions and Outlook}
\label{sec:conclusion}

In the present work, the characteristics of the higher order moments of the particle multiplicity are analysed by means of  PLSM. To this end, we first started with studying the critical phenomena which are accessible by this QCD--like model. Then, we estimated some thermodynamic quantities and compared them with different first--principle lattice QCD simulations. The first four non--normalized moments are investigated in thermal and density medium. Also, we studied the thermal and dense dependence of the first four moments normalized to the temperature and the chemical potential. Their characteristic non--monotonic behavior in thermal and dense medium is estimated. The possible fluctuations associated with the second order moment \cite{Schaefer:2008hk,Karsch2009a} are utilized in mapping out the chiral phase--transition. The peak is characterized by a set of $T$-- and $\mu$--values. 

We studied dependence of the chiral condensates and the order parameters on temperature and chemical potential.  We estimated the dependence of the potential on the temperature, chemical potential and minimization parameter. The latter assures a minimum potential, as well. Apparently, this depends on temperature and chemical potential. Additionally, we have other four parameters, $\sigma_x$, $\sigma_y$, $\phi $ and $\phi^*$. Therefore, the minimization step should be repeated for at least one of these parameters, while the other ones should remain fixed, i.e. {\it global minimum of other parameters}. Repeating this process, we get each parameter as a function of temperature and chemical potential.

At vanishing chemical potential, the PLSM  pressure  is compared with two sets of lattice calculations. Although, the general temperature--dependence of the pressure is not absence, the comparison with the most recent ones is not convincing, especially above $T_c$. We also studied the thermal behavior of PLSM pressure at finite chemical potential. The PLSM trace anomaly gets maximum around the critical temperature. This qualitatively agrees with the behaviour observed in lattice QCD calculations. Again, a systematic improvement to the PLSM configurations would be needed to reproduce the most recent lattice data.  With new configurations, we simply mean replacing the Polyakov--loop potential with the gluonic sector of quasi--particle model \cite{qpm}.

At fixed chemical potentials, the thermal dependence of the higher order moments becomes non--monotonic with increasing the orders and the chemical potentials, as well. At fixed temperatures, the dependence on the medium density is also non--monotonic. We conclude that PLSM seems to be sensitive to the fluctuations  accompanying the different order moments. It is apparent that increasing temperature weakens the fluctuations. Thus, it is conjectured that the phase transition could be indicated by such fluctuations.  In the present work, we can implement two types of scan in order to map out the boundary of the chiral phase transition. The thermal dependence of the higher order moments is practically related to a scan in $T$-direction, while the dependence on the density refers to a scan in $\mu$-direction.

When confronting PLSM to the freeze--out parameters deduced from lattice QCD calculations and thermal models, we find that the latter lay slightly above PLSM results. On the other hand, the freeze--out parameters at large chemical potential are currently questioned. Therefore, it is too early to draw any conclusion about the whole chiral phase diagram. But, we may understand that the freeze--out boundary and chiral phase--transition are almost non--distinguishable, especially at small chemical potential. We intend to confront new configurations of  PLSM to the most recent lattice calculations and even experimental measurements.


\appendix

\section{Derivatives of first order moment}
\label{appnd:1}
{\footnotesize
The first derivative of the total potential Eq. (\ref{Pr}) with respect to the chemical potential $\mu$ gives the number or multiplicity density. This quantity estimates the number of quarks and antiquarks contributions, Eq. (\ref{quarks}). 
\bea
\label{quarks}
n_{\psi \bar{\psi}}&=&\frac{T}{\pi ^{2}}\int d^{3}p \lbrace  -3 ( 2 \sqrt{4p^2 +g^2 \sigma _{x}^{2}} +g^2 \sigma _{x}\,\sigma _{x}^{'}+\, e^{\frac{2 \mu +\sqrt{4p^2 +g^2 \sigma _{x}^{2}}}{2T}} (2 \phi \lbrace (2 \sqrt{4p^2 +g^2 \sigma _{x}^{2}} +g^2 \sigma _{x}\,\sigma _{x}^{'})+ \nn \\
  && +e^{\frac{2 \mu +\sqrt{4p^2 +g^2 \sigma _{x}^{2}}}{2T}}( \sigma _{y}\, 2 \sqrt{4p^2 +g^2 \sigma _{x}^{2}} +g^2 \sigma _{x}\,\sigma _{x}^{'})-\, 2T\, \sqrt{4p^2 +g^2 \sigma _{x}^{2}}\left(\phi ^{'}+\sigma _{y} ^{'}  e^{\frac{2 \mu +\sqrt{4 p^2 +g^2 \sigma _{x}^{2}}}{2T}} \right)\rbrace ) \nn \\
 && \frac{1}{2T\, \sqrt{4p^2 +g^2 \sigma _{x}^{2}}\left(1+e^{\frac{3(2 \mu +\sqrt{4p^2 +g^2 \sigma _{x}^{2}})}{2T}}+3 \phi e^{\frac{2 \mu +\sqrt{4p^2 +g^2 \sigma _{x}^{2}}}{2T}}+3 \sigma _{y} e^{\frac{2 \mu +\sqrt{4p^2 +g^2 \sigma _{x}^{2}}}{2T}}\right)} +\nn \\ 
&&
  3\,e^{\mu /T}  ( 2\,e^{2\,\mu /T} \sqrt{2p^2 +g^2 \sigma _{y}^{2}} + \sqrt{2}\,e^{2\,\mu /T}\,g^2 \sigma _{y}\,\sigma _{y}^{'}+\, e^{\frac{\sqrt{4p^2 +2\,g^2 \sigma _{y}^{2}}}{T}} \nn \\ 
  &&( \phi \lbrace (2 \sqrt{2p^2 +g^2 \sigma _{y}^{2}} -\sqrt{2}\,g^2 \sigma _{y}\,\sigma _{y}^{'})+ 2e^{\frac{2 \mu +\sqrt{4p^2 +g^2 \sigma _{y}^{2}}}{2T}} \phi^{*}(2 \sqrt{2p^2 +g^2 \sigma _{y}^{2}} -\sqrt{2}\,g^2 \sigma _{y}\,\sigma _{y}^{'}) \nn  \\
  && {}-\, 2T\, \sqrt{2p^2 +g^2 \sigma _{y}^{2}}\phi ^{'} \, e^{\frac{\sqrt{4p^2 +2\,g^2 \sigma _{y}^{2}}}{T}}+
  2\,T \phi^{*'}  e^{\frac{2 \mu +\sqrt{4 p^2 +2g^2 \sigma _{y}^{2}}}{2T}} \sqrt{2p^2 +g^2 \sigma _{y}^{2}} \rbrace ) \nn  \\ 
   &&\frac{1}{2T\, \sqrt{2p^2 +g^2 \sigma _{y}^{2}}\left(e^{\frac{3\mu}{T}}+e^{\frac{3(2 \mu +\sqrt{p^2 +\frac{1}{2} g^2 \sigma _{y}^{2}})}{T}}+3 \phi e^{\frac{ \mu +\sqrt{4p^2 +g^2 \sigma _{x}^{2}}}{T}}+3\phi ^{*} e^{\frac{4 \mu +\sqrt{4p^2 +2 g^2 \sigma _{y}^{2}}}{2T}}\right)}- \nn \\ 
   && 6 ( 2 \sqrt{4p^2 +g^2 \sigma _{x}^{2}} +g^2 \sigma _{x}\,\sigma _{x}^{'}+\, e^{\frac{2 \mu +\sqrt{4p^2 +g^2 \sigma _{x}^{2}}}{2T}} (2 \phi \lbrace (2 \sqrt{4p^2 +g^2 \sigma _{x}^{2}} +g^2 \sigma _{x}\,\sigma _{x}^{'})+  \nn  \\
   && {}+e^{\frac{2 \mu +\sqrt{4p^2 +g^2 \sigma _{x}^{2}}}{2T}} \phi ^{*}(\, 2 \sqrt{4p^2 +g^2 \sigma _{x}^{2}} +g^2 \sigma _{x}\,\sigma _{x}^{'})-\, 2T\, \sqrt{4p^2 +g^2 \sigma _{x}^{2}}\left(\phi ^{'}+\phi ^{*'}  e^{\frac{2 \mu +\sqrt{4 p^2 +g^2 \sigma _{x}^{2}}}{2T}} \right)\rbrace ) \nn \\
   &&\frac{1}{2T\, \sqrt{4p^2 +g^2 \sigma _{x}^{2}}\left(1+e^{\frac{3(2 \mu +\sqrt{4p^2 +g^2 \sigma _{x}^{2}})}{2T}}+3 \phi e^{\frac{2 \mu +\sqrt{4p^2 +g^2 \sigma _{x}^{2}}}{2T}}+3 \phi ^{*} e^{\frac{2 \mu +\sqrt{4p^2 +g^2 \sigma _{x}^{2}}}{2T}}\right)}+ \nn \\
    &&
6\,e^{\mu /T}  ( 2\,e^{2\,\mu /T} \sqrt{2p^2 +g^2 \sigma _{x}^{2}} -e^{2\,\mu /T}\,g^2 \sigma _{x}\,\sigma _{x}^{'}+\, e^{\frac{\sqrt{4p^2 +2\,g^2 \sigma _{x}^{2}}}{T}}  \nn \\ 
&&( \phi \lbrace (2 \sqrt{4\,p^2 +g^2 \sigma _{x}^{2}} -\,g^2 \sigma _{x}\,\sigma _{x}^{'})+ 2\,e^{\frac{2 \mu +\sqrt{4p^2 +g^2 \sigma _{x}^{2}}}{2T}} \phi^{*}(2 \sqrt{4\,p^2 +g^2 \sigma _{x}^{2}} -\,g^2 \sigma _{x}\,\sigma _{x}^{'}) \nn  \\
&& {}+\, 2T\, \sqrt{2p^2 +g^2 \sigma _{x}^{2}}\phi ^{'} \, e^{\frac{\sqrt{4p^2 +2\,g^2 \sigma _{x}^{2}}}{T}}+
  2\,T \phi^{*'}  e^{\frac{2 \mu +\sqrt{4 p^2 +2g^2 \sigma _{x}^{2}}}{2T}} \sqrt{4p^2 +g^2 \sigma _{x}^{2}} \rbrace ) \nn  \\ 
&& \frac{1}{2T\, \sqrt{2p^2 +g^2 \sigma _{y}^{2}}\left(e^{\frac{3\mu}{T}}+e^{\frac{3(2 \mu +\sqrt{p^2 +\frac{1}{2} g^2 \sigma _{y}^{2}})}{T}}+3 \phi e^{\frac{ \mu +\sqrt{4p^2 +g^2 \sigma _{x}^{2}}}{T}}+3\phi ^{*} e^{\frac{4 \mu +\sqrt{4p^2 +2 g^2 \sigma _{y}^{2}}}{2T}}\right)}\rbrace
\eea
%
and the multiplicity density stemming from Polyakov-loop, Eq. (\ref{Polyakov1})
%
\bea
n_{loop}(\phi ,\phi ^{*},\phi ^{'},\phi ^{*'},T) &=& T^4 \left[ \frac{1}{2} b_{2}(T) \, \left(\phi \, \phi ^{'}+\, \phi ^{*}\, \phi ^{*'} \right)+\frac{1}{2} b_{3} \left(\phi ^{2}\, \phi ^{'} +\, \phi ^{*2}\,\phi ^{*'}\right)-\frac{1}{4}\,b_{4}\,\left(\phi ^{2}\,+\,\phi ^{*2}\right)\left(\phi \, \phi ^{'}+\phi ^{*}\, \phi ^{*'}\right)\right]. \hspace*{10mm} \label{Polyakov1}
\eea
}

\section{Derivatives of second order moment}
\label{appnd:2}
{\footnotesize
The  derivative of the multiplicity density  with respect to the chemical potential $\mu$ gives the susceptibility of the measurements. This consists of a linear combination of 
Polyakov-loop in Eq. (\ref{Polyakov}) and quarks and antiquarks contributions Eq. (\ref{quarks:sup}).
%
\bea
\chi _{loop}(\sigma _x,\sigma _x',\sigma _x'',\sigma _y,\sigma _y',\sigma _y'',T) &=& T^4 \lbrace   -\frac{1}{2}\, b_{2}(T)\left( \phi ^{'2} +\, \phi ^{*'2} +\, \phi \,\phi ^{''}+\, \phi ^{*}\phi ^{*''}\right)  \nn{} \\
&& -\frac{b_{3}}{2}\left(2 \,\phi \phi ^{'2}+2   \,\phi ^{*} \phi ^{*'2}+\,  \,\phi \phi ^{2} \,\phi \phi ^{''}+\, \phi ^{*2}\phi ^{*''}\right)\, \nn \\
&& +\frac{b_{4}}{8}\left[ \left( 2\phi \, \phi ^{'}+\phi ^{*} \, \phi ^{*'}\right)^{2}+\left( \phi ^{2}+ \phi ^{*2}\right) \left(2 \phi ^{'2}+2 \phi ^{*'2}+2 \phi ^{'}\phi ^{''}+2 \phi ^{*'}\phi ^{*''}\right)\right]\rbrace  \hspace*{10mm}\label{Polyakov} \\
\chi _{\psi \bar{\psi}}&=& -\frac{T}{\pi ^2}\int d^3 P\left[A1+A2+A3+A4+A5+2(A6+A7+A8+A9)\right], \label{quarks:sup}
\end{eqnarray}
where
\bea
A1&=& 81( e^{\frac{\sqrt{2 g^2 \sigma _y^2+4 P^2}+2 \mu }{2 T}}+2
   \sqrt{g^2 \sigma _y^2+2 P^2}+\sqrt{2} g^2 \sigma _y \sigma
   _y' +e^{\frac{\sqrt{2 g^2 \sigma _y^2+4 P^2}+2 \mu }{2 T}}  \\  \nn  && {} (2 \phi \left(2 \sqrt{2 P^2 + g^2 \sigma _y^2}+\sqrt{2} g^2 \sigma _y \sigma _y'\right)+\phi ^* \left(2 \sqrt{g^2 \sigma _y^2+2 P^2}+\sqrt{2} g^2
   \sigma _y \sigma _y'\right)  e^{\frac{\sqrt{2 g^2 \sigma
   _y^2+4 P^2}+2 \mu }{2 T}}-  \\  \nn  && {} 2 T \sqrt{g^2 \sigma _y^2+2 P^2} \left( \phi ^{* '}
   e^{\frac{\sqrt{2 g^2 \sigma _y^2+4 P^2}+2 \mu }{2 T}}+\Phi '\right)))^2  \\ \nn {}&&
   \frac{1}{\left(3 e^{-\frac{\sqrt{\frac{1}{2} g^2 \sigma _y^2+P^2}+\mu
   }{T}} \left(\Phi  e^{-\frac{\sqrt{\frac{1}{2} g^2 \sigma
   _y^2+P^2}+\mu }{T}}+\Phi ^*\right)+e^{-\frac{3
   \left(\sqrt{\frac{1}{2} g^2 \sigma _y^2+P^2}+\mu
   \right)}{T}}+1\right)^2} +\\
   A2&=&
(9\,e^{2\mu /T}  ( 2 e^{\frac{2 \mu }{T}} \sqrt{g^2 \sigma _y^2+2 P^2}+2 T \phi ' \sqrt{g^2 \sigma _y^2+2 P^2} e^{\frac{\sqrt{2 g^2 \sigma _y^2+4 P^2}}{T}}\phi  e^{\frac{\sqrt{2 g^2 \sigma _y^2+4 P^2}}{T}} \left(2 \sqrt{g^2 \sigma _y^2+2 P^2}-\sqrt{2} g^2 \sigma _y \sigma _y'\right)-\\ \nn {} && \sqrt{2} g^2 e^{\frac{2 \mu }{T}} \sigma _y \sigma _y' +  2 T \phi ^{*'} \sqrt{g^2 \sigma _y^2+2 P^2} e^{\frac{\sqrt{2 g^2 \sigma _y^2+4 P^2}+2 \mu }{2 T}}+2
   \phi ^* e^{\frac{\sqrt{2 g^2 \sigma _y^2+4 P^2}+2 \mu }{2 T}}[2 \sqrt{g^2 \sigma _y^2+2 P^2}- \sqrt{2} g^2 \sigma _y(\mu ) \sigma _y'])^2
 \\ \nn {}&& \frac{1}{4 T^2 \left(g^2 \sigma _y^2+2 P^2\right)  \left(3 \phi ^* e^{\frac{\sqrt{2 g^2 \sigma _y^2+4 P^2}+4 \mu }{2 T}}+3  \phi  e^{\frac{\sqrt{2 g^2 \sigma _y^2+4 P^2}+\mu}{T}}+e^{\frac{3 \sqrt{\frac{1}{2} g^2 \sigma _y^2+P^2}}{T}}+e^{\frac{3 \mu }{T}}\right)^2},
\eea
\bea 
 A3&=&(
\frac{3 \left(\frac{g^2 \sigma _y \left(\sigma _y\right)'}{\sqrt{2 g^2 \sigma _y^2+4 P^2}}+1\right){}^2 e^{-\frac{\sqrt{\frac{1}{2} g^2 \sigma _y^2+P^2}+ \mu }{T}} \left(\phi  e^{-\frac{\sqrt{\frac{1}{2} g^2 \sigma _y^2+P^2}+\mu }{T}}+\phi ^*\right)}{T^2}+\frac{9 \left(\frac{g^2 \sigma _y \left(\sigma _y\right)'}{\sqrt{2 g^2
   \sigma _y^2+4 P^2}}+1\right){}^2 e^{-\frac{3 \left(\sqrt{\frac{1}{2}
   g^2 \sigma _y^2+P^2}+\mu \right)}{T}}}{T^2}-\\ \nn {} && \frac{3 g^2  e^{-\frac{\sqrt{2 g^2 \sigma _y^2+4 P^2}+2 \mu }{T}} \left(\sigma _y  \sigma _y '' (g^2 \sigma _y^2+2 P^2 +2 P^2
   \sigma _y ^{' 2}\right)  \left(\phi ^* e^{\frac{\sqrt{2 g^2 \sigma _y^2+4 P^2}+2 \mu }{2 T}}+\phi \right)}{\sqrt{2} T \left(g^2 \sigma _y^2+2 P^2\right)^{3/2}}  )\\ \nn {} && \frac{1}{3 e^{-\frac{\sqrt{\frac{1}{2} g^2 \sigma _y^2+P^2}+\mu }{T}} \left(\phi 
   e^{-\frac{\sqrt{\frac{1}{2} g^2 \sigma _y^2+P^2}+\mu }{T}}+\phi
   ^*\right)+e^{-\frac{3 \left(\sqrt{\frac{1}{2} g^2 \sigma _y^2+P^2}+\mu
   \right)}{T}}+1} \\
A4&=&[\frac{3 g^2 e^{-\frac{3 \left(\sqrt{2 g^2 \sigma _y^2+4 P^2}+2 \mu \right)}{2 T}} \left(\sigma _y \sigma _y'' \left(g^2 \sigma
   _y^2+2 P^2\right)+2 P^2 \left(\sigma _y'\right){}^2\right)}{\sqrt{2} T \left(g^2 \sigma _y^2+2 P^2\right){}^{3/2}}-\\ \nn && \frac{g^2 e^{-\frac{\sqrt{2 g^2 \sigma _y^2+4 P^2}+2 \mu }{T}} \left(\sigma _y \sigma _y''
   \left(g^2 \sigma _y^2+2 P^2\right)+2 P^2 \left(\sigma _y'\right){}^2\right) \left(\phi ^*
   e^{\frac{\sqrt{2 g^2 \sigma _y^2+4 P^2}+2 \mu }{2 T}}+\phi \right)}{\sqrt{2} \text{3T}
   \left(g^2 \sigma _y^2+2 P^2\right){}^{3/2}}\\ \nn &&
   \frac{e^{-\frac{\sqrt{2 g^2 \sigma _y^2+4 P^2}+2 \mu }{2 T}}}{2 T^2
   \left(g^2 \sigma _y^2+2 P^2\right){}^{3/2}} (+4 P^2 \sqrt{g^2 \sigma _y^2+2 P^2}\\ \nn && 
   2 T \left(g^2 \sigma _y^2+2 P^2\right) \left(T \sqrt{g^2 \sigma _y^2+2
   P^2} \left(\left(\Phi ^*\right)'' e^{\frac{\sqrt{2 g^2 \sigma _y^2+4
   P^2}+2 \mu }{2 T}}+\phi ''\right)-\phi ' \left(2 \sqrt{g^2 \sigma
   _y^2+2 P^2}+\sqrt{2} g^2 \sigma _y \sigma _y'\right)\right)+\\ \nn &&\phi 
   \left(g^2 \left(\sigma _y^2 \sqrt{g^2 \sigma _y^2+2 P^2} \left(g^2
   \left(\sigma _y'\right){}^2+2\right)+\sqrt{2} g^2 \sigma _y^3 \left(2
   \sigma _y'-T \sigma _y''\right)+2 \sqrt{2} P^2 \sigma _y \left(2
   \sigma _y'-T \sigma _y''\right)-2 \sqrt{2} P^2 T \left(\sigma
   _y'\right){}^2\right)\right))]
   \\ \nn {} && \frac{1}{3 e^{-\frac{\sqrt{\frac{1}{2} g^2 \sigma _y^2+P^2}+\mu }{T}} \left(\phi 
   e^{-\frac{\sqrt{\frac{1}{2} g^2 \sigma _y^2+P^2}+\mu }{T}}+\phi
   ^*\right)+e^{-\frac{3 \left(\sqrt{\frac{1}{2} g^2 \sigma _y^2+P^2}+\mu
   \right)}{T}}+1}
\eea  

\bea
A5&=&[\frac{3 \left(\sqrt{2} g^2 \sigma _y \sigma _y'-2 \sqrt{g^2 \sigma _y^2+2 P^2}\right){}^2
   e^{\frac{2 \mu -3 \sqrt{2 g^2 \sigma _y^2+4 P^2}}{2 T}} \left(\Phi ^* e^{\frac{\sqrt{2 g^2
   \sigma _y^2+4 P^2}+2 \mu }{2 T}}+\Phi  e^{\frac{\sqrt{2 g^2 \sigma _y^2+4 P^2}}{T}}+3
   e^{\frac{2 \mu }{T}}\right)}{4 T^2 \left(g^2 \sigma _y^2+2 P^2\right)}\\ \nn &&
   -\frac{\left(\frac{g^2 \sigma _y \sigma _y'}{\sqrt{2 g^2 \sigma _y^2+4 P^2}}-1\right) e^{-\frac{\sqrt{\frac{1}{2} g^2 \sigma  _y^2+P^2}-\mu }{T}} \left(\left(\Phi ^*\right)' e^{-\frac{\sqrt{\frac{1}{2} g^2 \sigma _y^2+P^2}-\mu }{T}}-\frac{\Phi ^*(\mu ) \left(\frac{g^2 \sigma _y \sigma _y'}{\sqrt{2 g^2 \sigma _y^2+4 P^2}}-1\right) e^{-\frac{\sqrt{\frac{1}{2} g^2 \sigma _y^2+P^2}-\mu }{T}}}{T}+\Phi '\right)}{\text{T6}}\\ \nn &&
   -\frac{3 g^2 e^{\frac{2 \mu -3 \sqrt{2 g^2 \sigma _y^2+4 P^2}}{2 T}} \left(\sigma _y
   \sigma _y'' \left(g^2 \sigma _y^2+2 P^2\right)+2 P^2 \left(\sigma
   _y'\right){}^2\right) \left(\phi ^* e^{\frac{\sqrt{2 g^2 \sigma _y^2+4 P^2}+2 \mu
   }{2 T}}+\phi  e^{\frac{\sqrt{2 g^2 \sigma _y^2+4 P^2}}{T}}+e^{\frac{2 \mu
   }{T}}\right)}{\sqrt{2} T \left(g^2 \sigma _y^2+2 P^2\right){}^{3/2}}+
   \\ \nn &&
   3 e^{-\frac{\sqrt{\frac{1}{2} g^2 \sigma _y^2+P^2}-\mu }{T}}(\frac{\Phi ^* \left(\frac{g^2 \sigma _y \sigma _y'}{\sqrt{2 g^2 \sigma _y^2+4 P^2}}-1\right){}^2 e^{-\frac{\sqrt{\frac{1}{2} g^2 \sigma _y^2+P^2}-\mu }{T}}}{T^2}- 3 e^{-\frac{\sqrt{\frac{1}{2} g^2 \sigma _y^2+P^2}-\mu }{T}}\times \\ \nn &&[
  \frac{\phi ^* \left(\frac{g^2 \sigma _y \sigma _y'}{\sqrt{2 g^2 \sigma _y^2+4 P^2}}-1\right){}^2 e^{-\frac{\sqrt{\frac{1}{2} g^2 \sigma _y^2+P^2}-\mu }{T}}}{T^2}- \frac{g^2 \Phi ^* e^{-\frac{\sqrt{2 g^2 \sigma _y^2+4 P^2}-2 \mu }{2 T}} \left(\sigma _y \sigma _y'' \left(g^2 \sigma _y^2+2 P^2\right)+2 P^2 \left(\sigma _y'\right){}^2\right)}{\sqrt{2} T \left(g^2
   \sigma _y^2+2 P^2\right){}^{3/2}}-\\ \nn &&\frac{2 \left(\Phi ^*\right)' \left(\frac{g^2 \sigma _y
   \sigma _y'}{\sqrt{2 g^2 \sigma _y^2+4 P^2}}-1\right) e^{-\frac{\sqrt{\frac{1}{2} g^2 \sigma
   _y^2+P^2}-\mu }{T}}}{T}+\phi ^{*''} e^{-\frac{\sqrt{\frac{1}{2} g^2 \sigma
   _y^2+P^2}-\mu }{T}}+\phi '']]\times \\ \nn &&
   \frac{1}{3 e^{-\frac{\sqrt{\frac{1}{2} g^2 \sigma _y^2+P^2}-\mu }{T}} \left(\Phi ^*
   e^{-\frac{\sqrt{\frac{1}{2} g^2 \sigma _y^2+P^2}-\mu }{T}}+\Phi \right)+e^{-\frac{3
   \left(\sqrt{\frac{1}{2} g^2 \sigma _y^2+P^2}-\mu \right)}{T}}+1}]
   \eea
\begin{eqnarray}
A6&=&-[-\frac{3 e^{-\frac{3 \left(\sqrt{\frac{1}{4} g^2 \sigma _x{}^2+\text{p}^2}+\mu \right)}{T}} \left(\frac{g^2 \sigma _x\sigma _x'}{4 \sqrt{\frac{1}{4} g^2 \sigma _x{}^2+\text{p}^2}}+1\right)}{T} \\ \nn &&
-\frac{3 e^{-\frac{\sqrt{\frac{1}{4} g^2 \sigma _x{}^2+\text{p}^2}+\mu }{T}} \left(\frac{g^2 \sigma _x \sigma _x'}{4 \sqrt{\frac{1}{4} g^2 \sigma _x{}^2+\text{p}^2}}+1\right) \left(\phi e^{-\frac{\sqrt{\frac{1}{4} g^2 \sigma _x{}^2+\text{p}^2}+\mu }{T}}+\phi ^*(\mu )\right)}{T}\\ \nn &&
3 e^{-\frac{\sqrt{\frac{1}{4} g^2 \sigma _x{}^2+\text{p}^2}+\mu }{T}} \left(\left(\phi ^*\right)'+\phi ' e^{-\frac{\sqrt{\frac{1}{4} g^2 \sigma _x{}^2+\text{p}^2}+\mu }{T}}-\frac{\phi  e^{-\frac{\sqrt{\frac{1}{4} g^2 \sigma _x{}^2+\text{p}^2}+\mu }{T}} \left(\frac{g^2 \sigma _x \sigma _x'}{4 \sqrt{\frac{1}{4} g^2 \sigma _x{}^2+\text{p}^2}}+1\right)}{T}\right)]^2\\ \nn &&\times
\frac{1}{\left(3 e^{-\frac{\sqrt{\frac{1}{4} g^2 \sigma _x{}^2+\text{p}^2}+\mu }{T}} \left(\phi  e^{-\frac{\sqrt{\frac{1}{4} g^2 \sigma _x{}^2+\text{p}^2}+\mu }{T}}+\phi ^*\right)+e^{-\frac{3 \left(\sqrt{\frac{1}{4} g^2 \sigma _x{}^2+\text{p}^2}+\mu \right)}{T}}+1\right)^2}\\
A7&=-\,&[
-\frac{3 e^{-\frac{3 \left(\sqrt{\frac{1}{4} g^2 \sigma _x{}^2+\text{p}^2}-\mu \right)}{T}} \left(\frac{g^2 \sigma _x \sigma _x'}{4 \sqrt{\frac{1}{4} g^2 \sigma _x{}^2+\text{p}^2}}-1\right)}{T}\\ \nn &&
-\frac{3 e^{-\frac{\sqrt{\frac{1}{4} g^2 \sigma _x{}^2+\text{p}^2}-\mu }{T}} \left(\frac{g^2 \sigma _x \sigma _x'}{4 \sqrt{\frac{1}{4} g^2 \sigma _x{}^2+\text{p}^2}}-1\right) \left(\phi ^* e^{-\frac{\sqrt{\frac{1}{4} g^2 \sigma _x{}^2+\text{p}^2}-\mu }{T}}+\phi \right)}{T}\\ \nn &&+
3 e^{-\frac{\sqrt{\frac{1}{4} g^2 \sigma _x{}^2+\text{p}^2}-\mu }{T}} \left(\left(\phi ^*\right)' e^{-\frac{\sqrt{\frac{1}{4} g^2 \sigma _x{}^2+\text{p}^2}-\mu }{T}}-\frac{\phi ^* e^{-\frac{\sqrt{\frac{1}{4} g^2 \sigma _x {}^2+\text{p}^2}-\mu }{T}} \left(\frac{g^2 \sigma _x \sigma _x'}{4 \sqrt{\frac{1}{4} g^2 \sigma _x {}^2+\text{p}^2}}-1\right)}{T}+\phi ' \right)]^2 \\ \nn &&
\frac{1}{\left(3 e^{-\frac{\sqrt{\frac{1}{4} g^2 \sigma _x {}^2+\text{p}^2}-\mu }{T}} \left(\phi ^* e^{-\frac{\sqrt{\frac{1}{4} g^2 \sigma _x{}^2+\text{p}^2}-\mu }{T}}+\phi \right)+e^{-\frac{3 \left(\sqrt{\frac{1}{4} g^2 \sigma _x {}^2+\text{p}^2}-\mu \right)}{T}}+1\right)^2}
\end{eqnarray}
\begin{eqnarray}
A8&=&\frac{9 e^{-\frac{3 \left(\sqrt{\frac{1}{4} g^2 \sigma _x {}^2+\text{p}^2}+\mu \right)}{T}} \left(\frac{g^2 \sigma _x \sigma _x' }{4 \sqrt{\frac{1}{4} g^2 \sigma _x {}^2+\text{p}^2}}+1\right){}^2}{T^2}+\\ \nn &&
\frac{3 e^{-\frac{\sqrt{\frac{1}{4} g^2 \sigma _x {}^2+\text{p}^2}+\mu }{T}} \left(\frac{g^2 \sigma _x \sigma _x' }{4 \sqrt{\frac{1}{4} g^2 \sigma _x {}^2+\text{p}^2}}+1\right){}^2 \left(\phi  e^{-\frac{\sqrt{\frac{1}{4} g^2 \sigma _x{}^2+\text{p}^2}+\mu }{T}}+\phi ^*\right)}{T^2}+\\ \nn &&(
\frac{6 e^{-\frac{\sqrt{\frac{1}{4} g^2 \sigma _x{}^2+\text{p}^2}+\mu }{T}} \left(\frac{g^2 \sigma _x \sigma _x'}{4 \sqrt{\frac{1}{4} g^2 \sigma _x{}^2+\text{p}^2}}+1\right)}{T} \times \\ \nn &&
\left(\phi ^*\right)'+\phi ' e^{-\frac{\sqrt{\frac{1}{4} g^2 \sigma _x{}^2+\text{p}^2}+\mu }{T}}-\frac{\phi  e^{-\frac{\sqrt{\frac{1}{4} g^2 \sigma _x{}^2+\text{p}^2}+\mu }{T}} \left(\frac{g^2 \sigma _x \sigma _x'}{4 \sqrt{\frac{1}{4} g^2 \sigma _x{}^2+\text{p}^2}}+1\right)}{T}\\ \nn &&
-\frac{3 e^{-\frac{3 \left(\sqrt{\frac{1}{4} g^2 \sigma _x{}^2+\text{p}^2}+\mu \right)}{T}} \left(\frac{g^2 \sigma _x'{}^2}{4 \sqrt{\frac{1}{4} g^2 \sigma _x{}^2+\text{p}^2}}+\frac{g^2 \sigma _x \sigma _x''}{4 \sqrt{\frac{1}{4} g^2 \sigma _x{}^2+\text{p}^2}}-\frac{g^4 \sigma _x{}^2 \sigma _x'{}^2}{16 \left(\frac{1}{4} g^2 \sigma _x{}^2+\text{p}^2\right){}^{3/2}}\right)}{T}+\\ \nn && ( (
\frac{3 e^{-\frac{\sqrt{\frac{1}{4} g^2 \sigma _x{}^2+\text{p}^2}+\mu }{T}} \left(\phi  e^{-\frac{\sqrt{\frac{1}{4} g^2 \sigma _x{}^2+\text{p}^2}+\mu }{T}}+\phi ^* \right)}{T}) \times
\\ \nn &&
(\frac{g^2 \sigma _x'{}^2}{4 \sqrt{\frac{1}{4} g^2 \sigma _x{}^2+\text{p}^2}}+\frac{g^2 \sigma _x \sigma _x''}{4 \sqrt{\frac{1}{4} g^2 \sigma _x{}^2+\text{p}^2}}-\frac{g^4 \sigma _x{}^2 \sigma _x'{}^2}{16 \left(\frac{1}{4} g^2 \sigma _x{}^2+\text{p}^2\right){}^{3/2}}))+
\\ \nn &&
3 e^{-\frac{\sqrt{\frac{1}{4} g^2 \sigma _x{}^2+\text{p}^2}+\mu }{T}} \times (
-\frac{2 \phi ' e^{-\frac{\sqrt{\frac{1}{4} g^2 \sigma _x{}^2+\text{p}^2}+\mu }{T}} \left(\frac{g^2 \sigma _x \sigma _x'}{4 \sqrt{\frac{1}{4} g^2 \sigma _x{}^2+\text{p}^2}}+1\right)}{T} \\ \nn &&
+\frac{\phi e^{-\frac{\sqrt{\frac{1}{4} g^2 \sigma _x{}^2+\text{p}^2}+\mu }{T}} \left(\frac{g^2 \sigma _x \sigma _x'}{4 \sqrt{\frac{1}{4} g^2 \sigma _x{}^2+\text{p}^2}}+1\right){}^2}{T^2}+\phi '' e^{-\frac{\sqrt{\frac{1}{4} g^2 \sigma _x{}^2+\text{p}^2}+\mu }{T}}\\ \nn &&+
\left(\phi ^*\right)''-\frac{\phi  e^{-\frac{\sqrt{\frac{1}{4} g^2 \sigma _x{}^2+\text{p}^2}+\mu }{T}} \left(\frac{g^2 \sigma _x'{}^2}{4 \sqrt{\frac{1}{4} g^2 \sigma _x{}^2+\text{p}^2}}+\frac{g^2 \sigma _x \sigma _x''}{4 \sqrt{\frac{1}{4} g^2 \sigma _x{}^2+\text{p}^2}}-\frac{g^4 \sigma _x{}^2 \sigma _x'{}^2}{16 \left(\frac{1}{4} g^2 \sigma _x{}^2+\text{p}^2\right){}^{3/2}}\right)}{T})
\\ \nn && \times
\frac{1}{3 e^{-\frac{\sqrt{\frac{1}{4} g^2 \sigma _x{}^2+\text{p}^2}+\mu }{T}} \left(\phi e^{-\frac{\sqrt{\frac{1}{4} g^2 \sigma _x{}^2+\text{p}^2}+\mu }{T}}+\phi ^*\right)+e^{-\frac{3 \left(\sqrt{\frac{1}{4} g^2 \sigma _x{}^2+\text{p}^2}+\mu \right)}{T}}+1}
\end{eqnarray}

\begin{eqnarray}
A9&=[&(\frac{9 e^{-\frac{3 \left(\sqrt{\frac{1}{4} g^2 \sigma _x{}^2+\text{p}^2}-\mu \right)}{T}} \left(\frac{g^2 \sigma _x \sigma _x'}{4 \sqrt{\frac{1}{4} g^2 \sigma _x{}^2+\text{p}^2}}-1\right){}^2}{T^2}
\\ \nn && +
\frac{3 e^{-\frac{\sqrt{\frac{1}{4} g^2 \sigma _x{}^2+\text{p}^2}-\mu }{T}} \left(\frac{g^2 \sigma _x \sigma _x')}{4 \sqrt{\frac{1}{4} g^2 \sigma _x{}^2+\text{p}^2}}-1\right){}^2 \left(\phi ^* e^{-\frac{\sqrt{\frac{1}{4} g^2 \sigma _x{}^2+\text{p}^2}-\mu }{T}}+\phi \right)}{T^2}
\\ \nn &&+[(-\frac{e^{-\frac{\sqrt{\frac{1}{4} g^2 \sigma _x{}^2+\text{p}^2}-\mu }{T}} \left(\frac{g^2 \sigma _x \sigma _x'}{4 \sqrt{\frac{1}{4} g^2 \sigma _x{}^2+\text{p}^2}}-1\right)}{\text{T6}}) \\ \nn &&
(
\left(\phi ^*\right)' e^{-\frac{\sqrt{\frac{1}{4} g^2 \sigma _x{}^2+\text{p}^2}-\mu }{T}}-\frac{\phi ^* e^{-\frac{\sqrt{\frac{1}{4} g^2 \sigma _x{}^2+\text{p}^2}-\mu }{T}} \left(\frac{g^2 \sigma _x \sigma _x'}{4 \sqrt{\frac{1}{4} g^2 \sigma _x{}^2+\text{p}^2}}-1\right)}{T}+\phi ')]\,
\\ \nn && 
-\frac{3 e^{-\frac{3 \left(\sqrt{\frac{1}{4} g^2 \sigma _x{}^2+\text{p}^2}-\mu \right)}{T}} \left(\frac{g^2 \sigma _x'{}^2}{4 \sqrt{\frac{1}{4} g^2 \sigma _x{}^2+\text{p}^2}}+\frac{g^2 \sigma _x \sigma _x''}{4 \sqrt{\frac{1}{4} g^2 \sigma _x{}^2+\text{p}^2}}-\frac{g^4 \sigma _x{}^2 \sigma _x'{}^2}{16 \left(\frac{1}{4} g^2 \sigma _x{}^2+\text{p}^2\right){}^{3/2}}\right)}{T}
+\\ \nn && 
[(\frac{3 e^{-\frac{\sqrt{\frac{1}{4} g^2 \sigma _x{}^2+\text{p}^2}-\mu }{T}} \left(\phi ^* e^{-\frac{\sqrt{\frac{1}{4} g^2 \sigma _x{}^2+\text{p}^2}-\mu }{T}}+\phi \right)}{T})
\\ \nn && \times(
\frac{g^2 \sigma _x'{}^2}{4 \sqrt{\frac{1}{4} g^2 \sigma _x{}^2+\text{p}^2}}+\frac{g^2 \sigma _x \sigma _x''}{4 \sqrt{\frac{1}{4} g^2 \sigma _x{}^2+\text{p}^2}}-\frac{g^4 \sigma _x{}^2 \sigma _x'{}^2}{16 \left(\frac{1}{4} g^2 \sigma _x{}^2+\text{p}^2\right){}^{3/2}})]+
 \\ \nn &&
3 e^{-\frac{\sqrt{\frac{1}{4} g^2 \sigma _x{}^2+\text{p}^2}-\mu }{T}} \times ( 
\frac{\phi ^* e^{-\frac{\sqrt{\frac{1}{4} g^2 \sigma _x{}^2+\text{p}^2}-\mu }{T}} \left(\frac{g^2 \sigma _x \sigma _x'}{4 \sqrt{\frac{1}{4} g^2 \sigma _x{}^2+\text{p}^2}}-1\right){}^2}{T^2} + \\ \nn &&
\phi ''-\frac{2 \left(\phi ^*\right)' e^{-\frac{\sqrt{\frac{1}{4} g^2 \sigma _x{}^2+\text{p}^2}-\mu }{T}} \left(\frac{g^2 \sigma _x \sigma _x'}{4 \sqrt{\frac{1}{4} g^2 \sigma _x{}^2+\text{p}^2}}-1\right)}{T}
+
\left(\phi ^*\right)'' e^{-\frac{\sqrt{\frac{1}{4} g^2 \sigma _x{}^2+\text{p}^2}-\mu }{T}}- \\ \nn &&
\frac{\phi ^* e^{-\frac{\sqrt{\frac{1}{4} g^2 \sigma _x{}^2+\text{p}^2}-\mu }{T}} \left(\frac{g^2 \sigma _x'{}^2}{4 \sqrt{\frac{1}{4} g^2 \sigma _x{}^2+\text{p}^2}}+\frac{g^2 \sigma _x \sigma _x''}{4 \sqrt{\frac{1}{4} g^2 \sigma _x {}^2+\text{p}^2}}-\frac{g^4 \sigma _x {}^2 \sigma _x' {}^2}{16 \left(\frac{1}{4} g^2 \sigma _x {}^2+\text{p}^2\right){}^{3/2}}\right)}{T}) ] \\ \nn && \times 
\frac{1}{3 e^{-\frac{\sqrt{\frac{1}{4} g^2 \sigma _x {}^2+\text{p}^2}-\mu }{T}} \left(\phi ^* e^{-\frac{\sqrt{\frac{1}{4} g^2 \sigma _x {}^2+\text{p}^2}-\mu }{T}}+\phi \right)+e^{-\frac{3 \left(\sqrt{\frac{1}{4} g^2 \sigma _x {}^2+\text{p}^2}-\mu \right)}{T}}+1}
\end{eqnarray}


\end{document}